\documentclass[iop,preprint]{emulateapj}
\usepackage{graphicx}
\usepackage{amsmath}
\usepackage{natbib}

\begin{document}\sloppy

\slugcomment{submitted to ApJ}
\shortauthors{Harris et al.}

\title{The PIPER Survey: I. An Initial Look at 
the Intergalactic Globular Cluster Population in the Perseus Cluster}

\author{William E.~Harris}
\affiliation{Department of Physics \& Astronomy \\
McMaster University \\
Hamilton ON L8S 4M1, Canada}
\email{harris@physics.mcmaster.ca}

\author{Rachel A.~Brown}
\affiliation{Department of Physics \& Astronomy \\
McMaster University \\
Hamilton ON L8S 4M1, Canada}

\author{Patrick R. Durrell}
\affiliation{Department of Physics and Astronomy \\
	Youngstown State University \\
Youngstown OH 44555, USA}

\author{Aaron J.~Romanowsky}
\affiliation{Department of Physics \& Astronomy, San Jos\'e State University\\
One Washington Square, San Jos\'e CA 95192, USA \\
and University of California Observatories,
70 1156 High Street, Santa Cruz CA 95064, USA}

\author{John Blakeslee}
\affiliation{National Research Council of Canada \\
Herzberg Astronomy and Astrophysics Research Centre \\
Victoria BC, Canada}

\author{Jean Brodie}
\affiliation{University of California Observatories\\
1156 High Street, Santa Cruz CA 95064, USA}

\author{Steven Janssens}
\affiliation{Department of Astronomy and Astrophysics\\
University of Toronto, 50 St.~George St, Toronto ON M5S 3H4, Canada}

\author{Thorsten Lisker}
\affiliation{Astronomisches Rechen-Institut, Zentrum f\"ur Astronomie der  
Universit\"at Heidelberg, M\"onchhofstra{\ss}e 12-14, 69120 Heidelberg, Germany}

\author{Sakurako Okamoto}
\affiliation{Subaru Telescope, National Astronomical Observatory of Japan, 
650 North A’ohoku Place, Hilo, HI 96720, U.S.A.}
\affiliation{National Astronomical Observatory of Japan, Osawa 2-21-1, Mitaka, Tokyo, 181-8588, JAPAN}
\affiliation{The Graduate University for Advanced Studies, Osawa 2-21-1, Mitaka, Tokyo 181-8588, JAPAN}

\author{Carolin Wittmann}
\affiliation{Astronomisches Rechen-Institut, Zentrum f\"ur Astronomie der  
Universit\"at Heidelberg, M\"onchhofstra{\ss}e 12-14, 69120 Heidelberg, Germany}

\date{\today}

\clearpage

\begin{abstract}
	We describe the goals and first results of a Program for Imaging
	of the PERseus cluster of galaxies (PIPER).  The first phase of the program builds on 
	imaging of fields obtained with the Hubble Space Telescope (HST) ACS/WFC and
	WFC3/UVIS cameras.  Our PIPER target fields with HST include
	major early-type galaxies including the active central giant NGC 1275;
	known Ultra-Diffuse Galaxies; and the Intracluster Medium.  
	The resulting two-color photometry in F475W and F814W
	reaches deep enough to resolve and measure the globular cluster (GC) populations in
	the Perseus member galaxies.  Here we present  
	initial results for eight pairs of outer fields that confirm the presence 
	of Intergalactic GCs (IGCs) in fields as distant as 740 kpc from the Perseus center 
(40\% of the virial radius of the cluster).
	Roughly 90\% of these IGCs are identifiably blue (metal-poor) but there is a
	clear trace of a red (metal-rich) component as well, even at these very remote distances.
\end{abstract}


\section{Introduction}

Perseus (Abell 426) at $d=75$ Mpc offers a rich and fascinating laboratory for galaxy evolution,
but it has not yet gained the level of attention that has been given, for example, to Virgo or Coma.
Perseus has a velocity dispersion 
in the range $\sigma_v \simeq 1000 - 1300$ km s$^{-1}$,
among the highest of clusters in the local universe
\citep{kent_sargent1983,struble_rood1991,girardi1996,weinmann_etal2011}; for
comparison, the Coma cluster has $\sigma_v =$ 1100 km s$^{-1}$ \citep{colless_dunn1996}.
The total mass of Perseus is almost $10^{15} M_{\odot}$ \citep{girardi_etal1998,simionescu_etal2011}.
Within the bounds of the cluster is a vast halo of X-ray gas and dark matter with virial
radius $r_{200} = 1.8$ Mpc \citep{simionescu_etal2011}, 
comparable to the most gas-rich and populous galaxy
clusters known \citep[e.g.][]{zhao_etal2013,loewenstein1994,main_etal2017}.

Like Virgo and Coma, Perseus contains many large early-type galaxies, the most notable of which is 
the central supergiant NGC 1275 (= 3C48 = Perseus A), which sits at the
center of the X-ray gas and the dynamical center of the cluster.  
NGC 1275 is perhaps the most extreme case in the local
universe where we see the ongoing growth of a Brightest Cluster Galaxy (BCG)
complete with cooling flows and feedback, though globally its stellar mass
is dominated by an old, passive population typical of a giant elliptical.
Within NGC 1275 is a spectacular web of H$\alpha$ filaments
extending to more than 30 kpc from the galaxy center, which itself contains
$\sim 10^{11} M_{\odot}$ of molecular gas
and extended regions of star formation \citep[e.g.][]{fabian_etal2011,matsushita_etal2013,canning_etal2014}.

\begin{figure*}[t]
\vspace{-0.0cm}
\begin{center}
\includegraphics[width=1.0\textwidth]{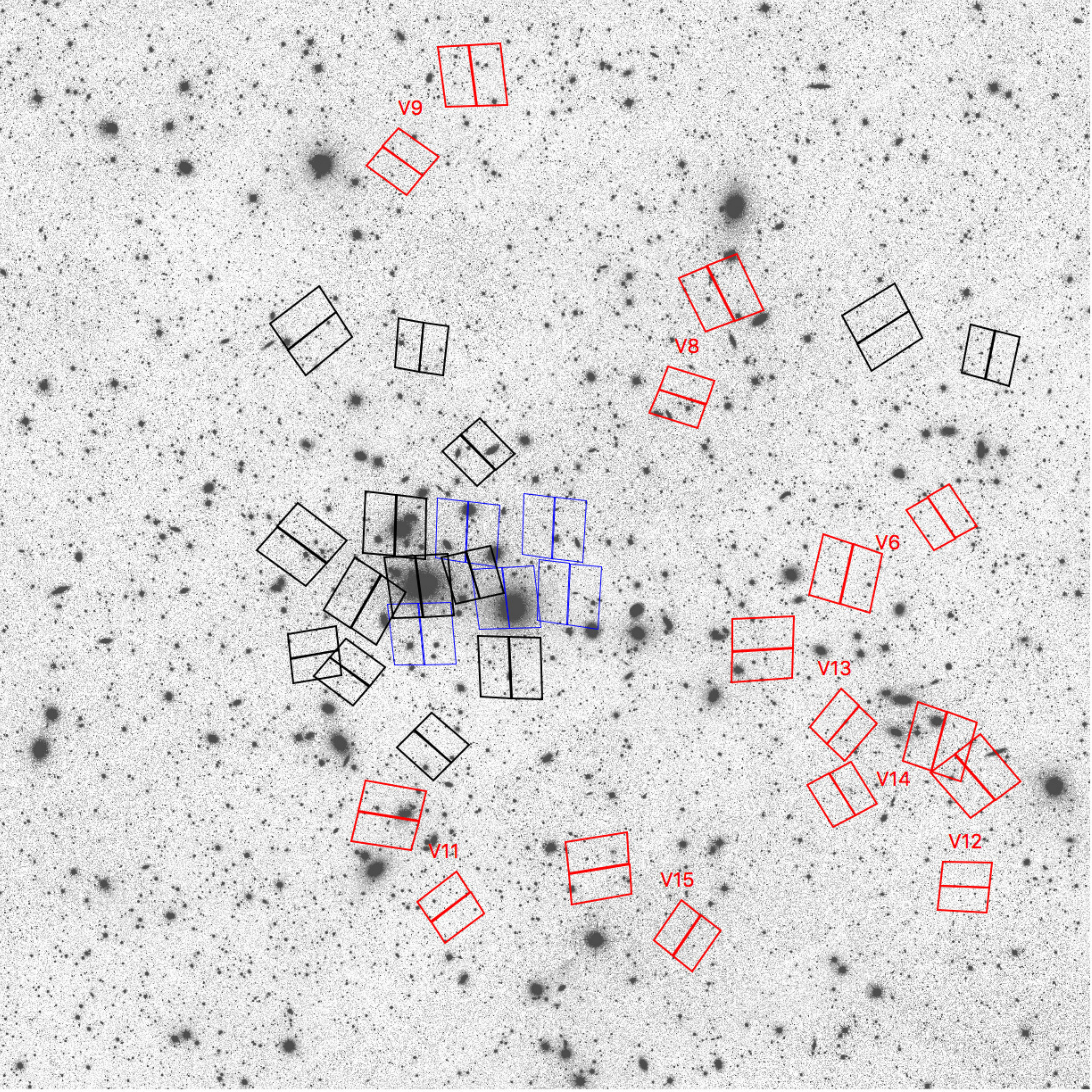}
\end{center}
\vspace{-0.2cm}
\caption{Image of the central region of the Perseus cluster generated
from an SDSS image stack in $r$.
North is at top, East at left and the field shown is $60' =$ 1300 kpc across.
The locations of the {\sl HST} pointings used in this study
are shown as pairs of rectangles (ACS the bigger, WFC3 the smaller in each pair).
NGC 1275, at the center of Perseus, is left of center.  The pointings specifically
discussed in this paper are marked in red and labelled with their Visit numbers
from Table 1.
The five thin blue single ACS fields near the center show
previous fields from the MAST archive, not used in the present paper but part of our
larger PIPER analysis to come.  }
\vspace{0.0cm}
\label{fig:fields}
\end{figure*}

\begin{table*}[t]
\begin{center}
\caption{\sc Target Fields}
\label{tab:coords}
\begin{tabular}{lrccrrrrl}
\tableline\tableline\\
\multicolumn{1}{l}{MAST Field Label} &
\multicolumn{1}{r}{Visit} &
\multicolumn{1}{c}{RA} &
\multicolumn{1}{c}{Dec}  &
\multicolumn{1}{r}{$R'$} &
\multicolumn{1}{r}{$A_I$} &
\multicolumn{1}{r}{$t_B,t_I$ (ACS)} &
\multicolumn{1}{r}{$t_B,t_I$ (WFC3)} &
\multicolumn{1}{l}{Target Galaxies} 
\\[2mm] \tableline\\
NGC1275-F1 & V1  & 03:19:49.6 & +41:30:36.0 & 0.29 & 0.245 & 2436,2325 & 2535,2535 & NGC 1275 \\
NGC1275-F2 & V2  & 03:20:03.5 & +41:29:39.4 & 3.06 & 0.245 & 2436,2325 & 2535,2535 & Perseus core \\
NGC1275-F3 & V3  & 03:19:54.5 & +41:33:54.5 & 3.42 & 0.247 & 2577,2429 & 2659,2772 & Perseus core \\
NGC1275-F4 & V4  & 03:19:23.1 & +41:26:09.3 & 6.53 & 0.242 & 2436,2325 & 2535,2535 & Perseus core \\
NGC1275-F5 & V5  & 03:20:23.8 & +41:32:43.1 & 6.97 & 0.249 & 2577,2429 & 2659,2772 & NGC 1278 \\
PERSEUS-UDG01 & V6  & 03:17:16.0 & +41:34:10.1 & 28.70 & 0.244 & 2424,2271 & 2533,2653 & WUDG 5,8,13,14 \\
PERSEUS-UDG02 & V7  & 03:17:00.3 & +41:42:58.9 & 33.74 & 0.227 & 2436,2105 & 2604,2640 & RUDG 5,6,84 \\
PERSEUS-UDG03 & V8  & 03:18:30.2 & +41:41:07.1 & 17.93 & 0.256 & 2436,2359 & 2604,2640 & WUDG 28,29,33,35,36,40,41, RUDG 25 \\
PERSEUS-UDG04 & V9  & 03:19:54.1 & +41:54:01.7 & 23.35 & 0.276 & 2436,2359 & 2604,2640 & RUDG 23,60 \\
PERSEUS-UDG05 & V10 & 03:19:50.2 & +41:43:19.3 & 12.63 & 0.252 & 2436,2539 & 2604,2640 & WUDG 83,84, RUDG 21,27 \\
PERSEUS-UDG06 & V11 & 03:19:39.9 & +41:13:07.6 & 17.64 & 0.228 & 2402,2271 & 2533,2653 & WUDG 79,80,88,89 \\
PERSEUS-UDG07 & V12 & 03:17:09.0 & +41:14:06.4 & 34.11 & 0.227 & 2424,2271 & 2533,2653 & WUDG 1,2,7, RUDG 15 \\
PERSEUS-UDG08 & V13 & 03:17:44.1 & +41:22:51.1 & 24.52 & 0.241 & 2424,2271 & 2533,2653 & WUDG 12,16,17,22 \\
PERSEUS-UDG09 & V14 & 03:17:45.5 & +41:19:30.2 & 25.55 & 0.238 & 2436,2359 & 2604,2640 & WUDG 4,6,7,18,19 \\
PERSEUS-UDG10 & V15 & 03:18:31.1 & +41:11:31.6 & 24.00 & 0.315 & 2424,2271 & 2533,2653 & WUDG 56,59, RUDG 16 \\
\\[2mm] \tableline
\end{tabular}
\end{center}
{\sc Note:} In the last column, WUDG identifiers are from \citet{wittmann_etal2017}
while the RUDG identifiers are additional UDGs from CFHT imaging (Romanowsky et al., in progress). \\
\vspace{0.4cm}
\end{table*}

In this paper, we describe a Program for Imaging of PERseus (called PIPER) 
currently underway that is
intended to attack four different program goals through the use of globular cluster (GC) populations.
\smallskip

\noindent (1) In recent years it has become clear that rich clusters of galaxies are hosts for
large numbers of Ultra-Diffuse Galaxies (UDGs), currently a focus of considerable interest:
recent observational work includes
\citet{vandokkum_etal2015a,vandokkum_etal2015b,mihos+2015,koda_etal2015,martinez-delgado_etal2016,roman_trujillo2017,papastergis_etal2017,amorisco_etal2018,janssens_etal2017}, 
now accompanied by a growing literature on theoretical modelling 
\citep[e.g.][]{rong+2017,chan+2018,jiang+2019,carleton+2019,martin+2019}.
Perseus is already known to hold many UDG candidates
\citep{wittmann_etal2017}, giving us the chance to improve our understanding
of their demographics.  Some UDGs in turn seem to have remarkably populous systems of GCs
relative to their low luminosities 
\citep{peng_lim2016,beasley_trujillo2016,vandokkum_etal2016,vandokkum_etal2018}, indicating that they
have massive dark halos and extremely high mass-to-light ratios \citep{harris_etal2017b},
while others are quite GC-poor
\citep{lim_etal2018,amorisco_etal2018}.  No highly consistent pattern has yet emerged,
and at least some of these extreme galaxies may violate the near-constant ratio of GC system mass
to halo (virial) mass obeyed by more luminous galaxies 
\citep{hudson_etal2014,durrell_etal2014,harris_etal2017b}.   Said differently,
for the lowest-luminosity galaxies including the UDGs, 
the scatter around either the $M_{GCS} - M_{vir}$ relation or
around the stellar-to-halo-mass relation (SHMR) may increase significantly;
for discussion of these issues, see the papers cited above and 
\citep{toloba_etal2018,elbadry_etal2018,amorisco_etal2018,lim_etal2018,forbes_etal2018,prole_etal2019,burkert_forbes2019}. 
\smallskip

\noindent (2) As a rich and dynamically active cluster, Perseus should also have
stellar Intra-Cluster Light (ICL) built from disrupted or stripped
member galaxies \citep[e.g.][]{burke_etal2012,ramos_etal2015,ramos_etal2018}.
In such clusters the ICL can make up typically 10--30\% of the total stellar mass.
Because the ICL is actively growing particularly since $z=1$, the sheer amount of ICL and its
degree of substructure (clumpiness and tidal streams) probe the
dynamical state of the entire cluster.  But the stellar ICL is extremely diffuse 
and difficult to map out with conventional surface-brightness photometry over 
a field as wide as the entire field of Perseus, whose 
virial radius is near $\sim 1.5^{\circ}$.  
For such a low signal, a more effective tracer is one
for which the background `noise' can be reduced to near-zero levels.
As will be shown below, GCs fit this bill beautifully: they are individually luminous, easy
to isolate, and can be reached by {\it {\sl HST}} for
galaxies out to $\sim 250$ Mpc or more \citep{harris_etal2016}.  

Substantial numbers of intergalactic globular clusters (IGCs) have been detected 
in other large clusters including Virgo \citep{lee_etal2010,durrell_etal2014,ko_etal2017,longobardi_etal2018}, 
Coma \citep{peng_etal2011,madrid_etal2018},
A1185 \citep{west_etal2011}, and A1689 \citep{alamo-martinez_blakeslee2017}.
Perhaps most importantly, multiband photometry also
automatically yields their \emph{metallicity distribution function} (MDF) since,
for old ($\gtrsim 3$ Gyr) clusters, GC color is
a monotonic function of [Fe/H] and is quite insensitive to age \citep[e.g.][]{peng_etal2006}.
The extremely faint, diffuse, integrated stellar light cannot provide this level of insight
at large distances from galaxy center.

\citet[][hereafter HM17]{harris_mulholland2017} carried out a preliminary investigation of
the GC populations within Perseus using MAST Archival ACS and WFPC2 fields.
They found evidence for IGCs,
albeit with large uncertainties.  In and near the Perseus core, any IGC component
can be easily confused with (or dominated by) the rich GC systems around the
giant early-type galaxies, while in the remoter outskirts of Perseus, their
conclusions relied on only a handful of shallower WFPC2 images.
\smallskip
 
\begin{figure}[t]
\vspace{-0.0cm}
\begin{center}
\includegraphics[width=0.5\textwidth]{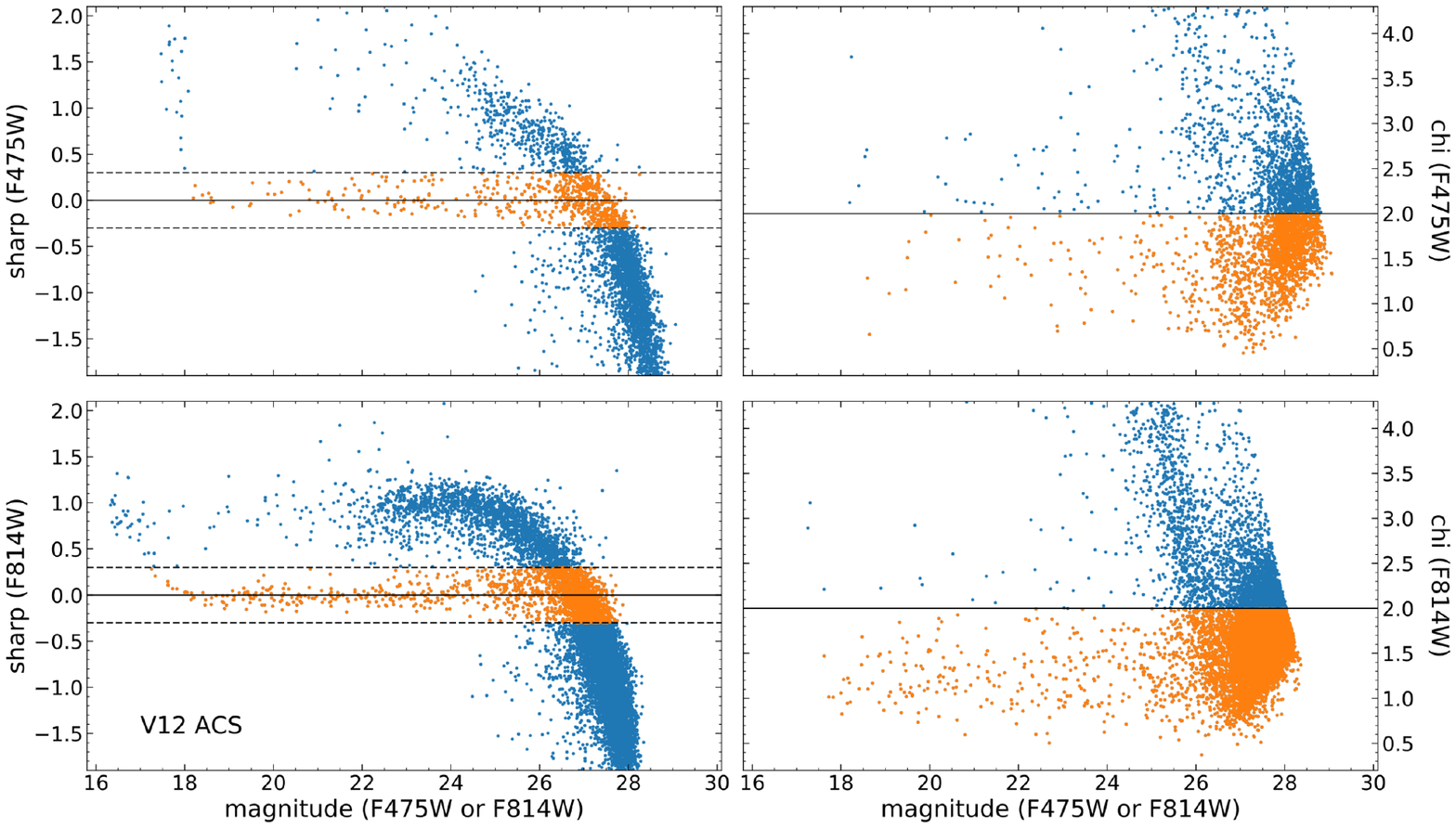}
\end{center}
\vspace{-6.2cm}
\caption{Illustration of the selection of GC candidates with the \emph{daophot} 
\emph{sharp} and \emph{chi} indices.  Magnitudes (either F475W or F814W) are
on the VEGAMAG system (see text).  The example shown is for the ACS V12 field.   
	For starlike (unresolved) objects \emph{sharp} is expected to be  $\simeq 0.0$.
Excluded objects categorized as nonstellar and rejected are shown with blue symbols;
accepted (near-starlike) objects with small \emph{sharp, chi} are in orange symbols.}
\vspace{0.0cm}
\label{fig:sharp1}
\end{figure}

\begin{figure}[t]
\vspace{-0.0cm}
\begin{center}
\includegraphics[width=0.5\textwidth]{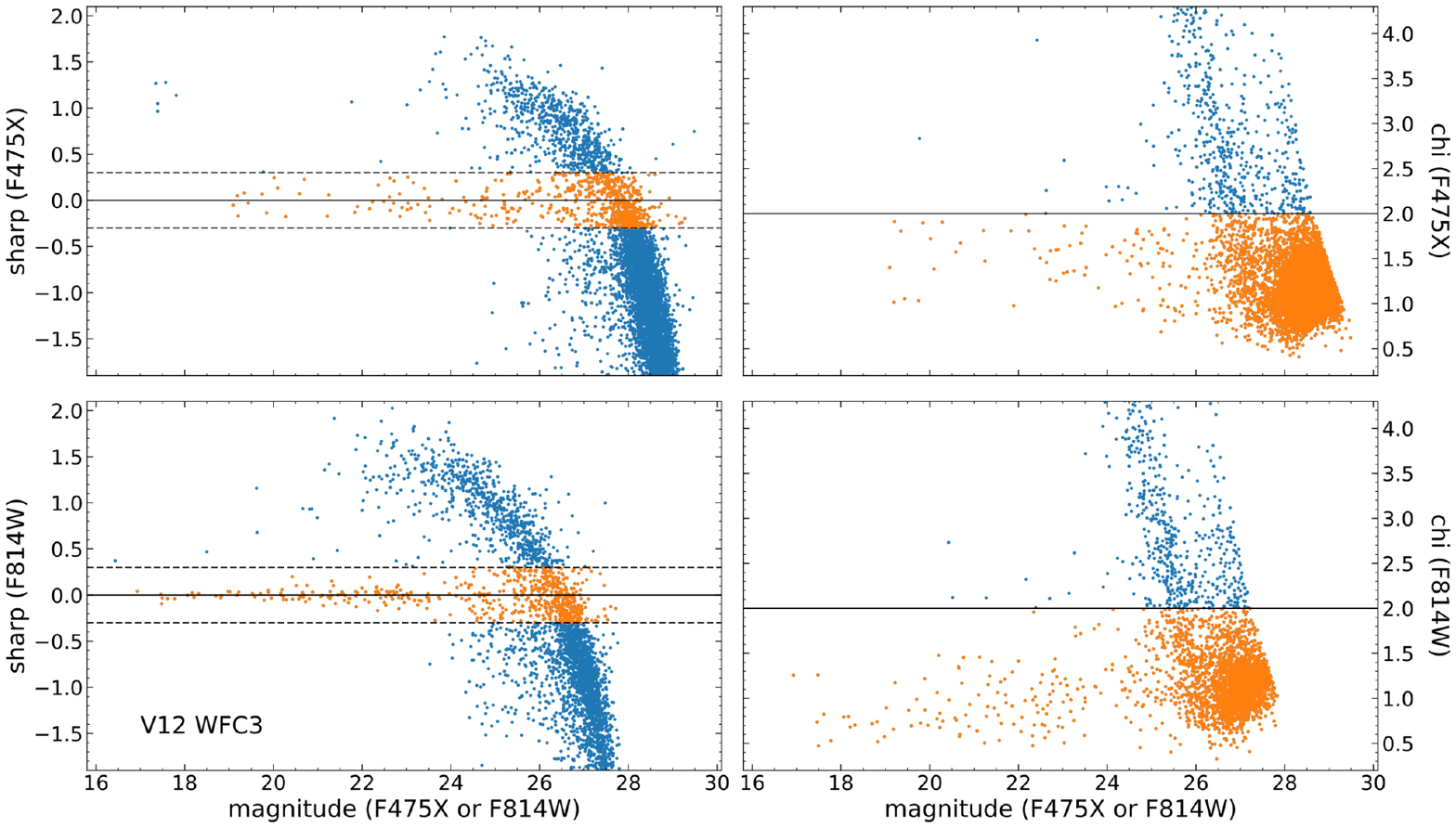}
\end{center}
\vspace{-6.2cm}
\caption{Selection of GC candidates with the \emph{daophot} 
\emph{sharp} and \emph{chi} indices, for the WFC3 V12 field.
Symbols are as in the previous figure.}
\vspace{0.0cm}
\label{fig:sharp2}
\end{figure}

\noindent (3) The Perseus core has several large early-type galaxies (ETGs), particularly NGC 1275,
NGC 1272, and NGC 1278.  Since their projected separations are $\lesssim 100$ kpc,
the total GC population in the core region will
be a mutually overlapping combination of these three major galaxies, plus some smaller Perseus 
members, plus the IGCs (which
should follow to first order the cluster potential well).  Given deep photometry with
appropriate field coverage of this central region,
a simultaneous solution for all these components can be performed.
NGC 1275 is uniquely interesting because
GC systems have never been explored in a galaxy with such extreme, high-activity conditions.
If, for example, its GCS turns out to have high specific frequency (number of GCs per
unit galaxy luminosity) as is
the usual case for BCGs \citep{harris_etal2017a}, it would support the view that
the current spectacular AGN and star-forming activity including young ($< 1$ Gyr)
star clusters in its inner 20 kpc
\citep{carlson_etal1998,canning_etal2010,canning_etal2014,lim_etal2019}  
is only an add-on to a dominant early formation epoch.
\smallskip

\noindent (4) At the opposite end of the
dwarf-galaxy structural scale from the UDGs are the Ultra-Compact Dwarfs
(UCDs).  Their characteristic luminosities $\gtrsim 10^7 L_{\odot}$ and
radii $r_h \gtrsim 10$ pc distinguish them from all
but the largest and most luminous GCs.  These dense stellar systems may be
a mixed population, either remnant nuclei of stripped dwarfs, or very massive
star clusters \citep[see][for recent discussion]{mieske_etal12,wittmann_etal2016,voggel_etal2018}.
At the Perseus distance, 
UCD candidates will be found as an automatic byproduct of our HST imaging data.
From the observed scaling relation between numbers of UCDs with
$M > 10^7 M_{\odot}$ and host cluster mass
\citep[e.g.][]{pfeffer2014,janssens_etal2017} Perseus should contain of order
200 UCDs. Some dozens of these are
already known  \citep{penny_etal2011,penny_etal2012}
and the results of our program are expected to yield 
enough UCDs in total to study the systematics of their spatial distributions
and mean metallicities.  

Lastly, we can expect to carry out an inventory of the 
M32-like compact ellipticals (cEs), a rare class of galaxy
now also emerging as a topic of systematic study \citep{janz_etal2016,martinovic_micic2017,zhang_bell2017,ferre-mateu_etal2018}.
\smallskip

For the following discussion, we adopt a distance $d = 75$ Mpc for Perseus, obtained from
the redshift 5207 km s$^{-1}$ corrected to the CBR frame (NED), 
and a Hubble constant $H_0 = 69.7$ km s$^{-1}$ Mpc$^{-1}$
\citep{hinshaw_etal2013}.  From NED the mean foreground extinction is
$A_I \simeq 0.246$, giving an apparent distance modulus $(m-M)_I = 34.62$.

The outline of this paper is as follows:  Section 2 describes the characteristics of our
imaging database and the photometric measurement techniques directed towards isolating
the GC candidates.  Section 3 presents 
the color-magnitude distribution of the GCs in each target field, and photometry
for a set of control fields well outside Perseus.  Section 4 presents in turn
the measurements of the GC 
effective radii, their luminosity distribution, color distribution, and radial
distribution within the Perseus cluster as a whole.
Section 5 concludes with a brief summary and goals for future work.

\section{Database and Observational Strategy}

Our imaging material consists of new fields observed with HST, along
with deep Subaru Hyper Suprime-Cam (HSC) imaging in $gri$ that covers a
$1^{\circ}$-diameter field, though to less depth and resolution.
In the present paper, we concentrate on the HST data for the outer
regions of the cluster.

Our new {\sl HST} imaging data for PIPER is from Cycle 25 program 15235
(PI Harris).  The ACS/WFC and WFC3/UVIS cameras are used in parallel to obtain
exposures of 15 pairs of fields scattered across the cluster; these pointings are
shown in Figure \ref{fig:fields} and listed in Table 1.  
In the Table, successive columns list the MAST Archive field
identifier; the Visit number; the field coordinates (J2000) for the Primary camera pointing
in each pair;
their projected distance $R$ in arcminutes from NGC 1275 (the Perseus center); mean foreground
extinction $A_I$ calculated from NED; the exposure times (in seconds) in the blue and red filters
from each of ACS and WFC3; and the UDG targets included in each pointing.  Exposure times
in parentheses are for Visits not yet executed at time of writing.
For Visits 1-5, the ACS camera is the Primary 
and WFC3 the Parallel; while for Visits 6-15, WFC3 is the Primary and ACS the Parallel.

The fields will be referred to below by their spacecraft Visit numbers V1--V15.
The first five Visits are the ones covering the Perseus core, while the remaining 10 
are the outer fields selected to cover the UDGs and, simultaneously,
the Intracluster Medium. Spacecraft orientations for the outer fields were chosen
to maximize the numbers of UDGs we could capture.  

The core of Perseus, at center-left in the figure,
is covered nearly completely out to $R \simeq 5'$ (100 kpc) 
when supplemented by five additional MAST Archive ACS/WFC fields taken from program GO-10201
(PI Conselice), also shown in the figure.  
As can be seen in Table 1, the field-to-field differences in foreground reddening are
modest (typically $\Delta A_I \lesssim 0.02$ mag) with the exception of the more
heavily reddened V15.

The schedule of spacecraft visits is spread out over a two-year period from
2017 Oct to 2019 October.  At time of writing, all image pairs except V2, V7, V10 have been obtained.
At a scale of 22 kpc/arcmin for a Perseus distance of 75 Mpc, V12 (the most remote pointing)
lies at a projected distance of 740 kpc from Perseus center (cf. Table 1), equivalent to 40\% of
the Perseus virial radius \citep{simionescu_etal2011}.

In the present paper, we discuss the outer fields V6--V15 (excluding V7, V10
not yet available) and specifically what they reveal about the population of IGCs.
These images are also used to define a consistent and homogeneous set of
measurement procedures for the entire set of HST data that will be discussed
in later papers in our series.
As will be shown below, Perseus clearly holds an IGC component, and the photometric
data are of sufficient quality to characterize their color (metallicity) distribution
and an initial estimate of their radial distribution within the cluster.

For most of our program goals we need to detect and characterize GCs within Perseus,
either in the target galaxies or distributed throughout the diffuse ICM.  
Typical GCs have half-light diameters 
$2r_h \simeq 6$ pc \citep[][2010 edition]{harris1996}, which at $d = 75$ Mpc 
translates to $0.014''$, almost an order of magnitude smaller than the natural $0.1''$ 
resolution of {\sl HST}.  This means that most GCs belonging to the Perseus galaxies will
be near-starlike (that is, unresolved) objects, which is a major advantage for
carrying out photometry and for isolating them from the field contaminants that
are dominated by faint, small background galaxies.  In addition,
GCs fall in a relatively narrow range of color index, permitting further rejection
of very blue or very red contaminants.  This combination of object morphology
and color gives us a very effective filtering procedure to isolate the GC population
with a low level of residual contamination (that is, a level where
the contaminating non-GC population makes only a second-order contribution
to the total, as will be seen in Section 3 below).

In all cases our program employs exposures in filters that maximize signal-to-noise
and can, if desired, be transformed to
standard $(B,I)$ for easy comparison with previous GC data in the literature.
The adopted filters are (F475W, F814W) for ACS/WFC; and (F475X, F814W) for WFC3/UVIS.
Though WFC3 F475X has rarely been used, it is significantly broader than F475W and still
transforms well to standard $B$ \citep{harris2018}.
For each field, either three or four dithered sub-exposures are taken in each filter to give a
final image effectively free of cosmic rays and other artifacts.
In what follows, we will refer to the exposure pairs as the $B$ and $I$ images,
although we present the results in the natural filter magnitudes.

The procedure for detection and photometry of the GC candidates is similar to the steps
outlined in, e.g., \citet{harris2009a,harris_etal2016} or HM17.
We start with the CTE-corrected, drizzled and combined \emph{*.drc} images provided by MAST.
First, the individual $B$ and $I$ exposures are registered and combined to create 
a master image in each filter.  In the \emph{pyraf} implementation of \emph{daophot},
\emph{daofind} is used independently on each filter to generate findlists of 
objects.  Then follows the normal sequence of 
\emph{phot} (through a $r=2$ px aperture), \emph{psf}, and \emph{ allstar}.  
The point-spread function (PSF) was defined empirically on each $B$ and $I$ image
individually from a combination of typically 30--50 moderately bright, uncrowded stars.
Tests were made to compare results adopting either a uniform PSF or one depending on position
$(x,y)$ in the image, but with entirely similar results; the uniform-PSF mode was adopted.
The \emph{allstar} magnitudes were then corrected to equivalent large-aperture magnitudes.
The appropriate photometric zeropoints for each filter and for the exact dates of the
exposures, as taken from the STScI webpage zeropoint calculators,
were then added to put the data onto the VEGAMAG system and in the natural filter magnitudes
(F475W, F814W for ACS, and F475X, F814W for WFC3).

Final rejection of distinguishably nonstellar objects was done with the \emph{sharp},
\emph{chi}, and \emph{err} parameters generated by \emph{allstar}, examples of which are shown 
in Figures \ref{fig:sharp1} and \ref{fig:sharp2}.  \emph{Sharp} is defined such that
starlike objects (ones closely matching the PSF) will appear near $\simeq 0$, 
and exclusion regions can then be defined to reject much smaller or more extended sources as
shown in the Figure.  Any objects with $|sharp| > 0.3$, $chi > 2$, or $err > 0.3$ on 
\emph{either} filter were rejected.\footnote{In practice, these three parameters are 
correlated.  That is, any object rejected because of high \emph{sharp} would also often 
have been rejected for high \emph{chi, err}. In general, however, the \emph{sharp} culling
was the first and most stringent step. By comparison, further culling by \emph{chi, err} rejected
only an additional few percent of the objects.}  
Extreme care was taken through the iterative PSF definition step
to ensure that the brighter, uncrowded stars in the image gave \emph{sharp} values very close
to zero. The internal scatter of the objects in the desired `starlike' sequence
at $sharp = 0$ differs slightly from one field and filter to another, so
fixed exclusion boundaries at $sharp = \pm0.3$ were chosen.  Setting the boundaries there
ensured that we included
virtually all objects in the starlike sequence in every image, while also excluding
the same numbers of contaminating field objects (see below).  

\begin{figure*}[t]
\vspace{-0.0cm}
\begin{center}
\includegraphics[width=0.75\textwidth]{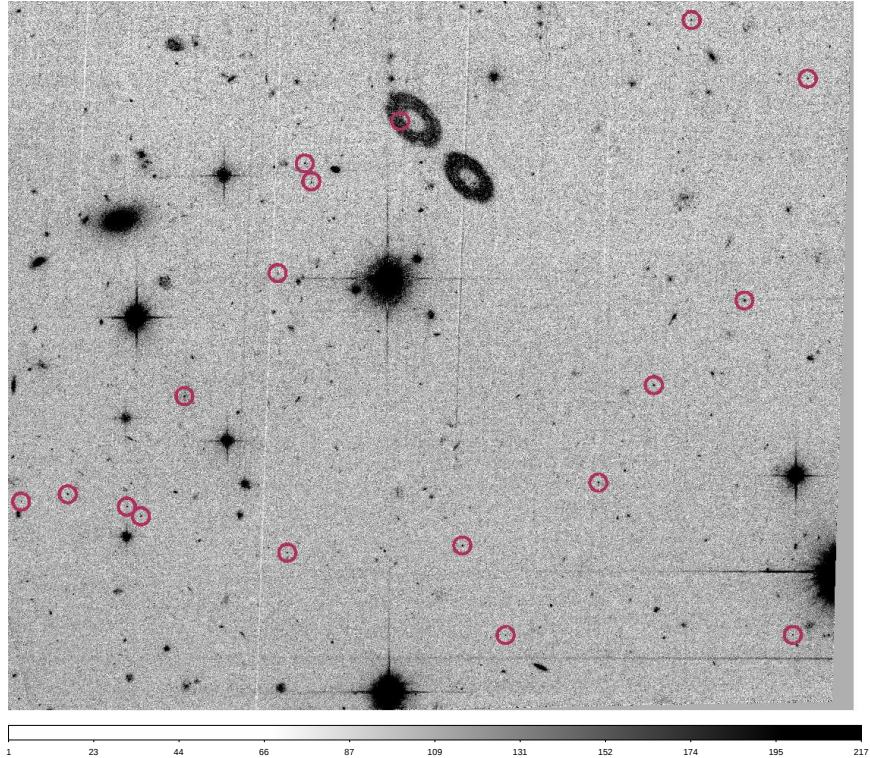}
\end{center}
\vspace{-0.2cm}
\caption{A section of the V15 field from the ACS camera, in F814W.  The segment shown is $100''$ 
	across.  The faint objects marked with the colored circles are GC
	candidates, which are starlike objects with $I < 26$ and colors in the range
$1.0 < (F475W-F814W)_0 < 2.4$.  The two oval rings above center are ghost images of adjacent
bright stars. }
\vspace{0.5cm}
\label{fig:v15view}
\end{figure*}

Finally, the culled object lists in each filter were matched to within $\Delta r = 2$ px in position,
further rejecting any objects not appearing in both filters.  
The entire combination of steps described above is effective not just 
at culling clearly nonstellar objects, but also at removing spurious 
detections near the CCD chip edges or within the bright cores of any Perseus member
galaxies falling within the fields.  

Saturation of starlike objects occurs for magnitudes F814W $\lesssim 18$, as can be seen
in Fig.~\ref{fig:sharp1}.  However, as will be seen below, the brightest GCs that are our
main targets lie safely fainter than this, at F814W $\gtrsim 22$.

\begin{figure*}[t]
\vspace{-0.0cm}
\begin{center}
\includegraphics[width=0.85\textwidth]{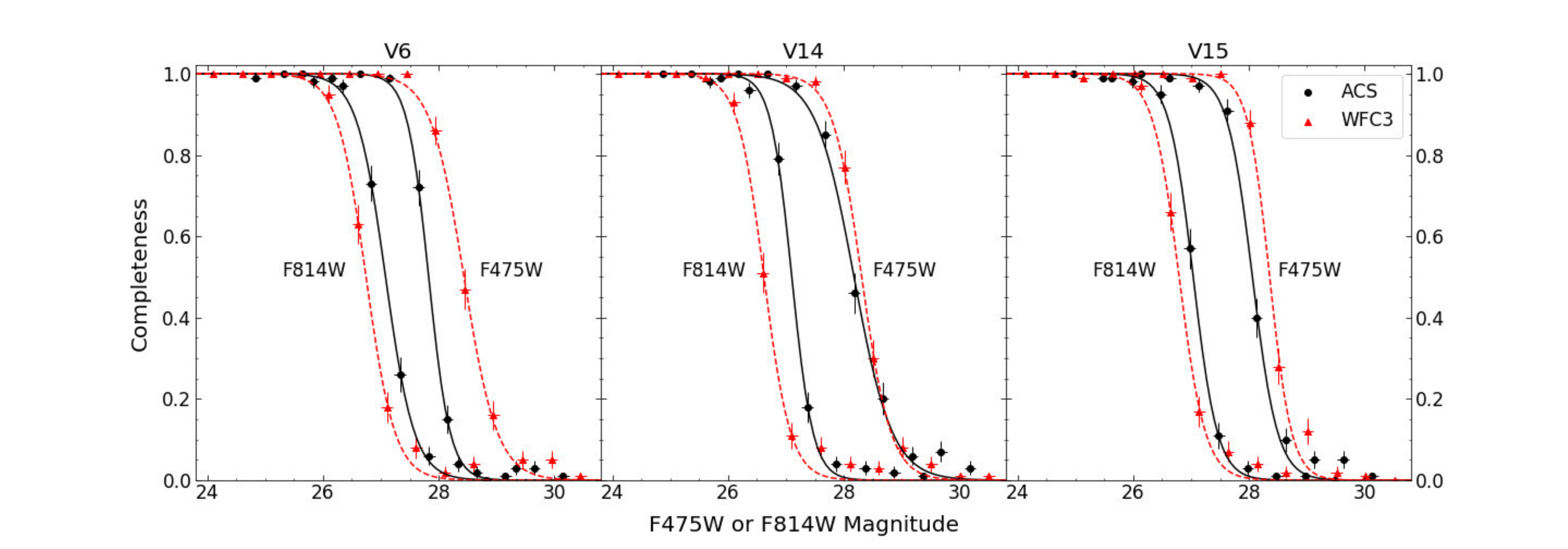}
\end{center}
\vspace{-0.2cm}
\caption{Photometric completeness tests for the ACS and WFC3 data.  The completeness ($f$) is the fraction of
	starlike objects at a given $B$ or $I$ magnitude that were detected, and is plotted versus 
	magnitude.  Black points and curves show the results for ACS, and red triangles 
and curves show the results for WFC3.
The completeness drops steeply from 100\% to near-zero over an interval of roughly a magnitude.}
\vspace{0.0cm}
\label{fig:f}
\end{figure*}

Photometric completeness of detection, and thus the effective limiting magnitudes of
our data, was evaluated by adding populations of artificial stars (scaled PSFs) 
into the images through \emph{daophot/addstar}.  The completeness ratio $f(m)$ is the number
of detected artificial stars at a given magnitude $m$ divided by the number of stars inserted at $m$.
All of these target fields are very sparsely populated and the sky intensity level is low and quite
uniform, as we illustrate in Figure \ref{fig:v15view}.  Although many small, faint background
galaxies can be seen scattered across the frame, the numbers of similarly faint unresolved
objects are quite small, and crowding is not an issue for any part of the data.
In addition, all the fields have nearly identical exposure times.
Under these almost ideal circumstances, the completeness fraction will
drop quite steeply from near-100\% to near-zero over a short magnitude range.
We model $f(m)$ with the simple two-parameter function \citep{harris_etal2016} 
$$	f = (1 + \mathrm{exp}(\alpha(m-m_0)))^{-1} $$
where $m_0$ is the magnitude
at which 50\% of the objects are detected and $\alpha$ measures the steepness of falloff
of the completeness curve.  
Examples of the completeness test results are shown in Figure \ref{fig:f},
while the best-fit parameters and their uncertainties are listed in Table \ref{tab:f}.
For the ACS data we find $m_0 \simeq 28.0 (F475W), 27.0 (F814W)$, while
for WFC3 (the more blue-sensitive of the two cameras) $m_0 \simeq 28.4 (F475X), 26.7 (F814W)$.  

\begin{table*}[t]
\begin{center}
\caption{\sc Photometric Completeness}
\label{tab:f}
\begin{tabular}{lrr}
\tableline\tableline\\
\multicolumn{1}{l}{Field} &
\multicolumn{1}{l}{$\alpha$} &
\multicolumn{1}{l}{$m_0$} \\
\\[2mm] \tableline\\
V6 ACS F475W & 5.4(0.2) & 27.83(0.01) \\
V6 ACS F814W & 4.0(0.2) & 27.08(0.02) \\
V6 WFC3 F475X & 3.5(0.2) & 28.44(0.02) \\
V6 WFC3 F814W & 4.0(0.3) & 26.74(0.02) \\
V14 ACS F475W & 3.0(0.3) & 28.18(0.03) \\
V14 ACS F814W & 5.5(0.3) & 27.10(0.01) \\
V14 WFC3 F475X & 4.0(0.2) & 28.31(0.01) \\
V14 WFC3 F814W & 4.4(0.5) & 26.62(0.02) \\
V15 ACS F475W & 4.6(0.5) & 28.06(0.02) \\
V15 ACS F814W & 4.8(0.2) & 27.04(0.01) \\
V15 WFC3 F475X & 5.3(0.6) & 28.36(0.03) \\
V15 WFC3 F814W & 4.4(0.6) & 26.79(0.02) \\
\\[2mm] \tableline
\end{tabular}
\end{center}
\vspace{0.4cm}
\end{table*}

\section{Detecting the IGCs}

Any GCs found in our (8+8) target fields are remote from the major Perseus
galaxies and therefore will be associated either with dwarf galaxies or the
even less luminous UDGs 
in those fields, or with the IntraCluster Medium.  Decades of previous work on GC populations
in other galaxies have established a remarkably consistent set of properties of GC systems,
including their luminosity distributions, color (metallicity) distributions, and GC scale sizes.
We look briefly at each of these properties in turn, and in doing so build up 
the case that the Perseus IGC population is clearly detected.

From analysis of a set of {\sl HST} Archival fields (some of which are marked in Fig.~\ref{fig:fields}),
HM17 found suggestive evidence for the presence of IGCs.  However, as noted above, their
preliminary findings relied in the end on a few WFPC2 fields with much shallower exposures than
the ones in our program, and on ACS fields located in
the core region near the giants NGC 1272 and 1275, making it difficult to isolate a
significant sample free of contamination from Perseus galaxies.  The data in our program
are better designed to find the intergalactic component, however much of it
there is.

\begin{figure*}[t]
\vspace{-0.0cm}
\begin{center}
\includegraphics[width=0.90\textwidth]{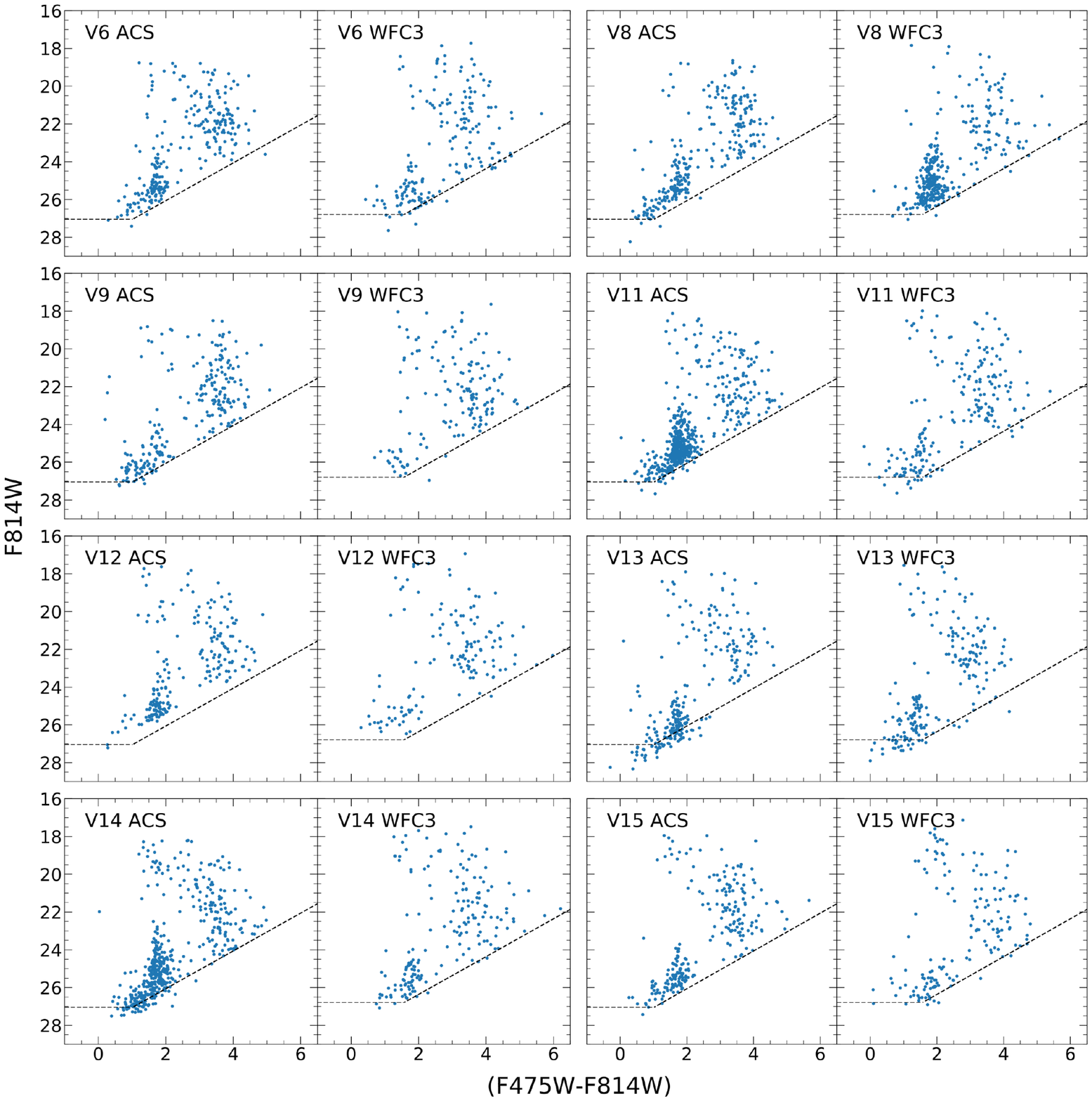}
\end{center}
\vspace{-4.6cm}
\caption{Color-magnitude data for starlike (unresolved) objects in eight individual pairs
of fields. The data are not de-reddened.  The dashed line indicates the average 50\% 
detection completeness level for the ACS fields as a rough guide to the limits of our
data.}  
\vspace{0.0cm}
\label{fig:cmd8}
\end{figure*}

\subsection{Color-Magnitude Data}

After the image matching and rejection of nonstellar objects as described above, the 
resulting color-magnitude diagrams (CMDs) for our eight pairs of fields are shown
in Figure \ref{fig:cmd8}.  Here, the data from both cameras have been put onto
a common photometric system, which we have adopted as (F475W, F814W)$_{ACS}$.
As noted above, the red filter for the WFC3 images was F814W, whose effective wavelength
is very similar to the equivalent filter for ACS and thus F814W(WFC3) $\simeq$ F814W(ACS),
which in turn are closely equivalent to the Johnson/Cousins I band
\citep{harris2018}.
The blue filter for the WFC3 images was F475X, chosen as noted above for its wider
bandpass.  Though it maximizes the blue limiting magnitude for a given exposure time,
it is significantly redder than F475W and thus needs transformation into F475W$_{ACS}$. 
From the equations in \cite{sirianni_etal2005} and \cite{harris2018}, we obtain
$$ F475W_{ACS} = F475X_{WFC3} + 0.16 (F475X-F814W) $$
which has been applied to all the WFC3 measurements.

\begin{figure*}[t]
\vspace{-0.0cm}
\begin{center}
\includegraphics[width=0.85\textwidth]{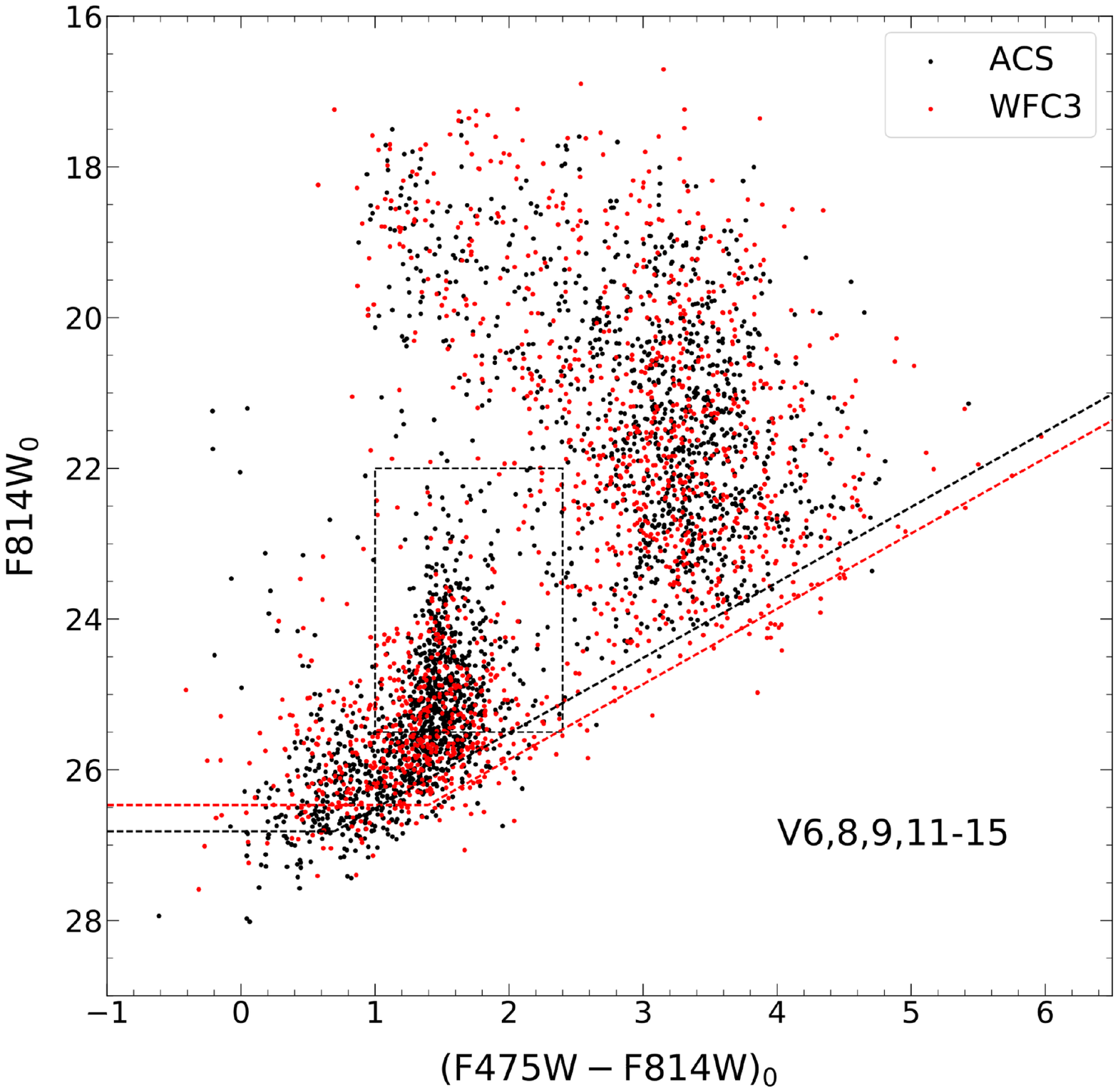}
\end{center}
\vspace{-5.6cm}
\caption{Composite CMD for all eight visits discussed in the present paper.  Black points are
the ACS images combined, red points are the WFC3 images. Objects falling within any
of the circles shown in the next Figure are excluded to remove any GCs belonging
to the dwarf galaxies in the fields.  The average 50\% completeness thresholds for each camera, established from the
completeness tests described in the text, are shown as the dashed lines (black for ACS,
red for WFC3). The box outlined in the black thin-dashed line marks
the GC candidate region defined in the text.}
\vspace{0.0cm}
\label{fig:cmdall}
\end{figure*}

\begin{figure*}[t]
\vspace{-0.0cm}
\begin{center}
\includegraphics[width=0.9\textwidth]{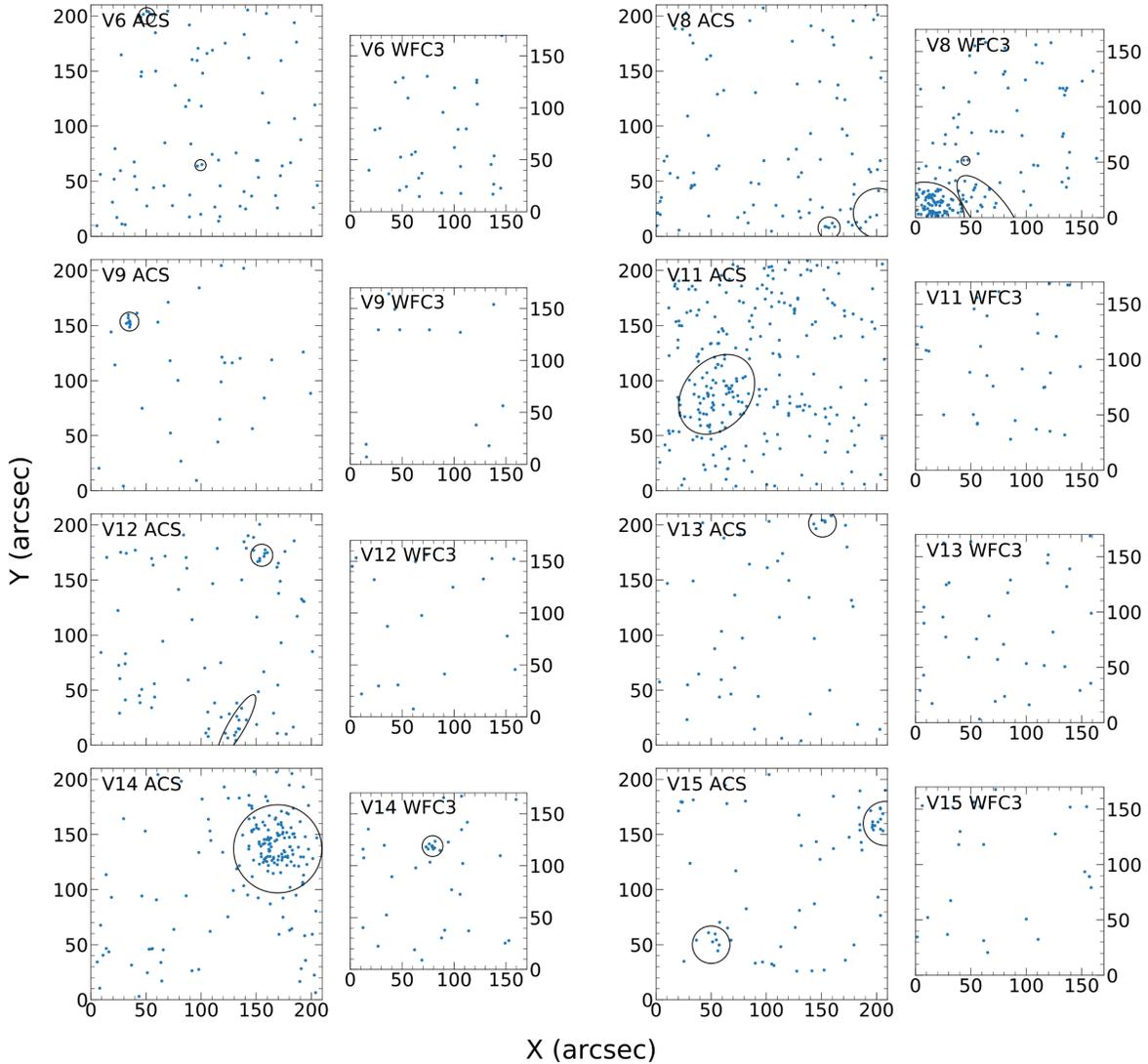}
\end{center}
\vspace{-5.2cm}
\caption{Locations of objects in each field that fall within the color and magnitude
	range for normal globular clusters ($1.0 < (F475W-F814W)_0 < 2.0$, $22.5 < F814W < 25.5$). 
	For each pair of panels, the left shows the ACS camera, and the right shows the WFC3 camera.
	Black circles and ellipses indicate the boundaries of small galaxies that have
detectable GC populations of their own; objects inside these regions are excluded from
the composite CMD shown in Fig.~\ref{fig:cmdall}. }
\vspace{0.0cm}
\label{fig:xyall}
\end{figure*}

In each field of Fig.~\ref{fig:cmd8} a distinct population of objects can be seen at intermediate color
(F475W-F814W) $\simeq 1.8/$ and magnitude range F814W $\gtrsim 23$, near the expected values for GCs.  
A composite CMD including data from all 16 individual fields is shown in Figure \ref{fig:cmdall},
where now each field has been individually dereddened so that the data can be strictly matched
together.  The CMD in Fig.~\ref{fig:cmdall} does, however, exclude
a few small concentrations of points around identifiable small galaxies in the target fields.
These exclusion regions are shown in Figure \ref{fig:xyall}. 
The distribution of these GC candidates across the CMD
is quite insensitive to the exact placement
of the spatial boundaries in Fig.~\ref{fig:xyall} around the dwarf galaxies, because of the well known property
of GCs in dwarfs to be almost entirely low-metallicity ones \citep{peng_etal2011}.
The combined CMD represents a total area of 150 arcmin$^2$. 

In this combined diagram, the GC candidate sequence now stands out more clearly.  The large
swath of points crossing the upper right of the CMD is dominated by foreground-star
contamination (see below), but fortunately this region is well separated from the target GCs.
Taking into account the
rapidly increasing scatter in color for $I > 26$ and the completeness limits in each filter, 
we isolate a ``GC candidate'' region generously defined by
the box $22.0 < F814W_0 < 25.5$, $1.0 < (F475W-F814W)_0 < 2.4$.  Within this box the photometry is also
highly complete ($>$90\%).  The boundaries of this region correspond to absolute magnitudes and
intrinsic colors $-12.4 < M_I < -8.9$ and (transformed color range) $1.2 < (B-I)_0 < 2.9$. 
For comparison, the GCs in a sample of Brightest Cluster Galaxies \citep{harris2009a}
populate the range $M_I > -12$ and $1.4 \lesssim (B-I)_0 \lesssim 2.3$, further indicating
that we are seeing a population of GCs.  
As will be seen below, the candidate GCs almost all fall in a somewhat narrower color
range $1.0 < (F475W-F814W)_0 < 2.0$.  However, in anticipation of later papers in this series, which will concentrate
on the GC populations around the Perseus giant ETGs that have much larger numbers of
more metal-rich GCs that may extend redder, we keep a slightly more generous color
boundary at the red end.

\begin{figure*}[t]
\vspace{-0.0cm}
\begin{center}
\includegraphics[width=0.55\textwidth]{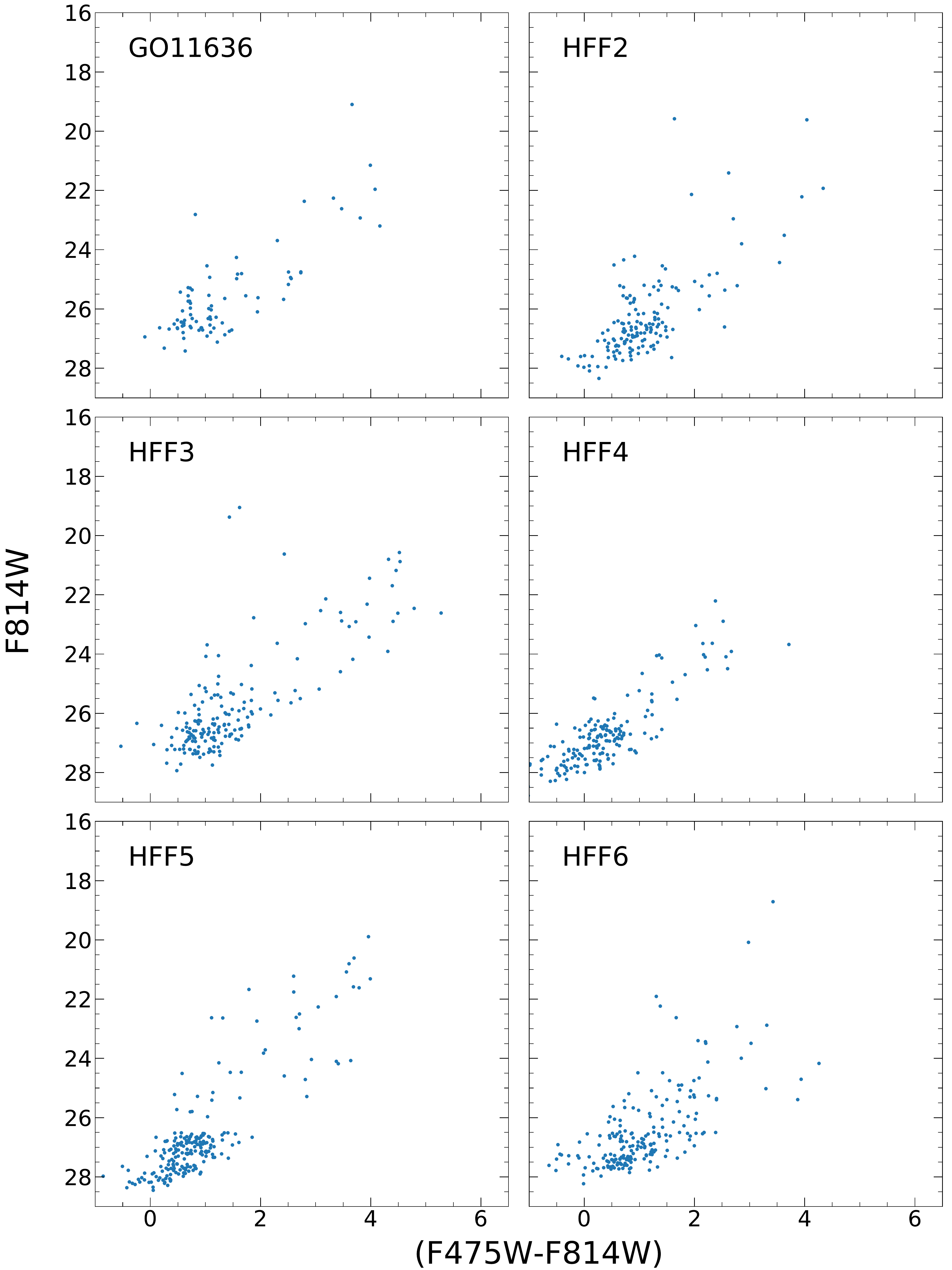}
\end{center}
\vspace{-0.0cm}
\caption{CMDs for the six HST/ACS control fields listed in Table \ref{tab:back}.
Photometric measurement and object selection were carried out with exactly the same procedures
as for the Perseus target fields. Compare with Fig.~\ref{fig:cmd8}.}
\vspace{0.0cm}
\label{fig:cmdback_tile}
\end{figure*}

\begin{figure*}[t]
\vspace{-0.0cm}
\begin{center}
\includegraphics[width=0.85\textwidth]{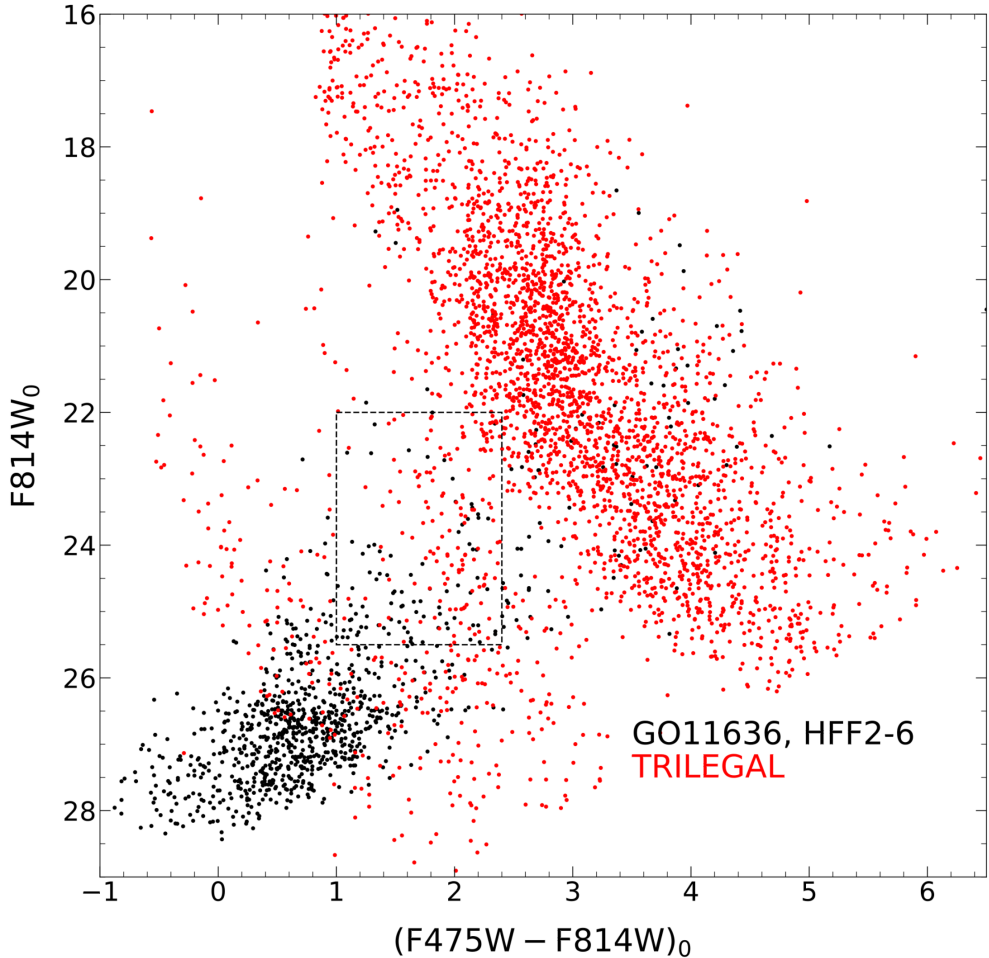}
\end{center}
\vspace{-5.2cm}
\caption{Combined CMD for expected field contamination.  Black dots are from the six control
	fields listed in Table \ref{tab:back}, while red dots are a sample of the predicted
	population of Milky Way foreground stars from the TRILEGAL model and described
in the text.
	The box marked out by the dashed line indicates the
GC-candidate region for Perseus; compare with Figure \ref{fig:cmdall}.}
\vspace{0.0cm}
\label{fig:cmdback}
\end{figure*}

\begin{figure}[t]
\vspace{-2.0cm}
\begin{center}
\includegraphics[width=0.5\textwidth]{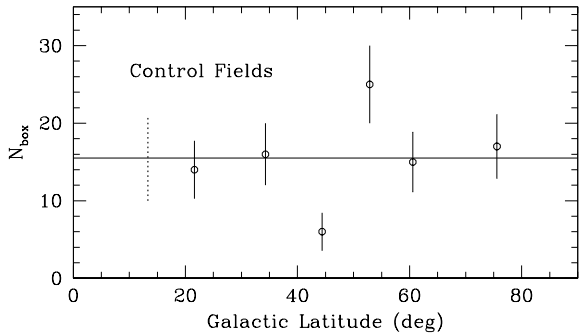}
\end{center}
\vspace{-0.6cm}
\caption{Number of measured objects falling within the GC candidate box, in each of the
six control fields, plotted versus absolute Galactic latitude $|b|$.
Errorbars assume $N^{1/2}$ Poisson statistics.
The horizontal line at $N_{box} = 15.5$ indicates the mean value over
the six fields.
The latitude of the Perseus cluster ($b = -13.3^o$) is marked with the vertical
dotted line at left.}
\vspace{0.0cm}
\label{fig:nbox}
\end{figure}

\begin{figure*}[t]
\vspace{-0.0cm}
\begin{center}
\includegraphics[width=0.5\textwidth]{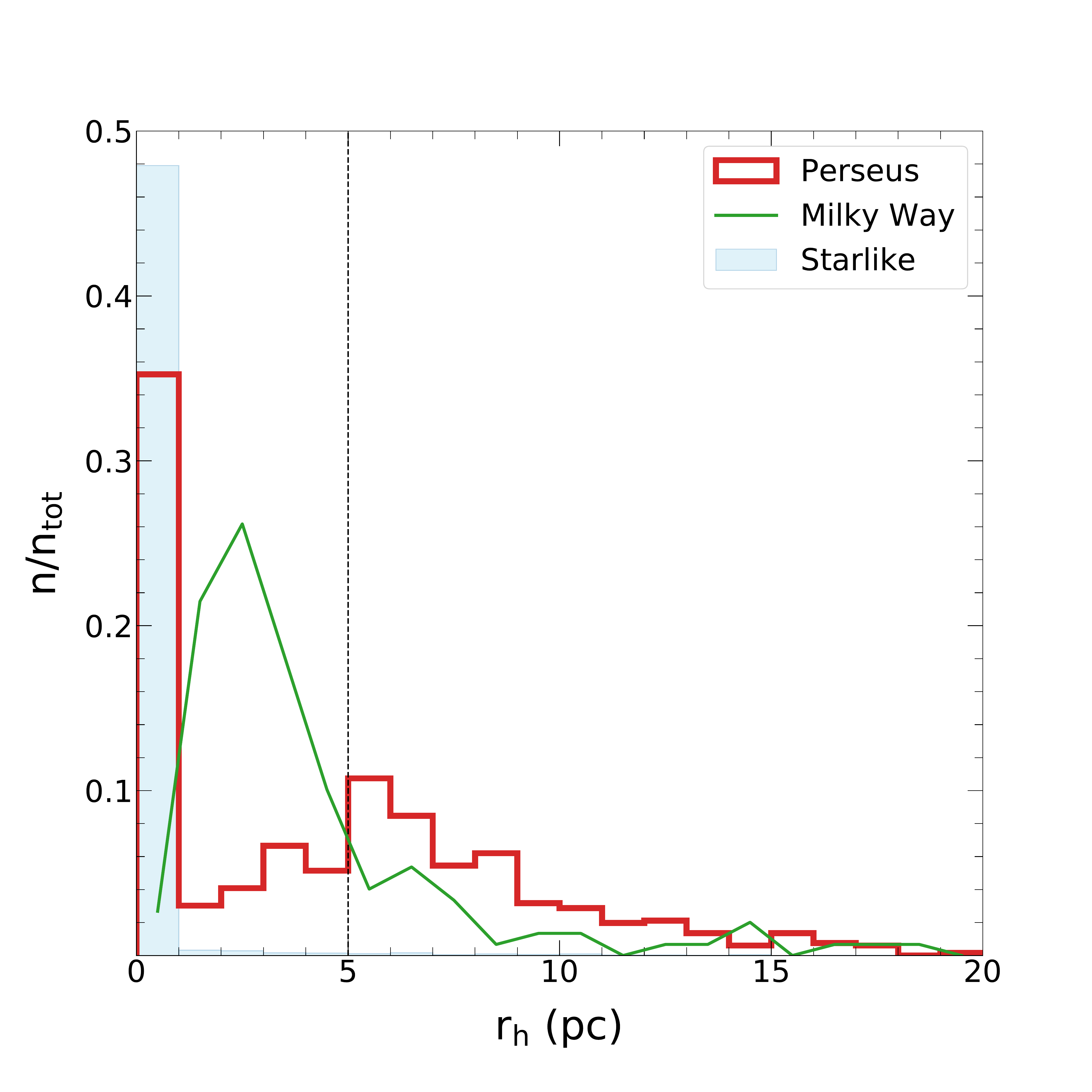}
\end{center}
\vspace{-0.2cm}
\caption{Histogram of intrinsic half-light radius ($r_h$ or $r_{eff}$) as deduced from the
	ISHAPE fitting; $r_h$ is shown in parsec units, plotted in 1-pc bins.  The Perseus objects in the 
	GC candidate list are shown as the red solid line; simulated starlike
	objects in the same magnitude range are shown as the cyan shaded histogram; and Milky Way GCs
as the solid green line. The vertical line at
$r_h \simeq 5$ pc indicates a conservative resolution limit above which $r_h$
should be measurable at the Perseus distance (see text).}
\vspace{0.9cm}
\label{fig:histo}
\end{figure*}

\begin{table*}[t]
\begin{center}
\caption{\sc Control Fields}
\label{tab:back}
\begin{tabular}{llrrccccc}
\tableline\tableline\\
\multicolumn{1}{l}{Field} &
\multicolumn{1}{l}{Identifier} &
\multicolumn{1}{l}{RA} &
\multicolumn{1}{l}{Dec}  &
\multicolumn{1}{l}{$\ell$}  &
\multicolumn{1}{l}{$b$}  &
\multicolumn{1}{c}{$A_I$} &
\multicolumn{1}{c}{$t_B$(sec)} &
\multicolumn{1}{c}{$t_I$(sec)} 
\\[2mm] \tableline\\
HFF2 & MACSJ0416.1-2403 & 04:16:33.1 & -24:06:48.7 & 327.3 & +34.3 & 0.134 & 3858 & 3855 \\
HFF3 & MACSJ0717.5+3745 & 07:17:17.0 & +37:49:47.3 & 180.5 & +21.6 & 0.104 & 3858 & 3855 \\
HFF4 & MACSJ1149.5+2223 & 11:49:40.5 & +22:18:02.3 & 230.5 & +75.6 & 0.039 & 3855 & 3855 \\
HFF5 & Abell S1063      & 22:49:17.7 & -44:32:43.8 & 349.4 & -60.6 & 0.021 & 3855 & 3855 \\
HFF6 & Abell 370        & 02:40:13.4 & -01:37:32.8 & 173.7 & -52.9 & 0.055 & 3858 & 3855 \\
SS22A-PAR & GO11636     & 22:17:14.5 & +00:14:12.3 &  63.8 & -44.3 & 0.103 & 3000 & 3000 \\
\\[2mm] \tableline
\end{tabular}
\end{center}
\vspace{0.4cm}
\end{table*}

\subsection{Background Contamination}

The fraction of this GC candidate sample that consists of actual Perseus GCs 
depends on the residual level of field contamination in the photometry.  This contamination
consists of 
foreground stars and very faint, small background galaxies that managed
to pass through the culling steps described above.  Ideally we would like to assess the
contamination level from background control fields near but outside Perseus, taken with ACS or 
WFC3 and with $B,I$ filters and similar exposure times to our program fields. However, we have found
no such ideal material in the MAST archive.
Instead, as a preliminary measure we use data
for six different fields listed in Table \ref{tab:back} that do have the right combinations
of filters and exposure times, but which lie in widely different parts of the sky.
These include five Hubble Frontier Fields 
\citep{lotz_etal2017} and one from program GO-11636.\footnote{One other HFF field, targeted
on Abell 2744, is not suitable for our purposes because its Parallel image is heavily
contaminated by galaxies in this very extensive cluster, and its own IGCs.}
All of these
are Parallel exposures with the ACS/WFC camera, and in all six cases they 
fall on `blank' fields free of any major galaxies or clusters of galaxies.  
We have subselected individual exposures from each program
adding up to total exposures similar to but a bit longer than our Perseus target fields, so that
they will have limiting magnitudes slightly deeper than our program fields.   
In Table \ref{tab:back}, we list the field coordinates, exposure times in the blue and red filters, and foreground extinctions $A_I$.

\begin{figure}[t]
\vspace{-0.0cm}
\begin{center}
\includegraphics[width=0.5\textwidth]{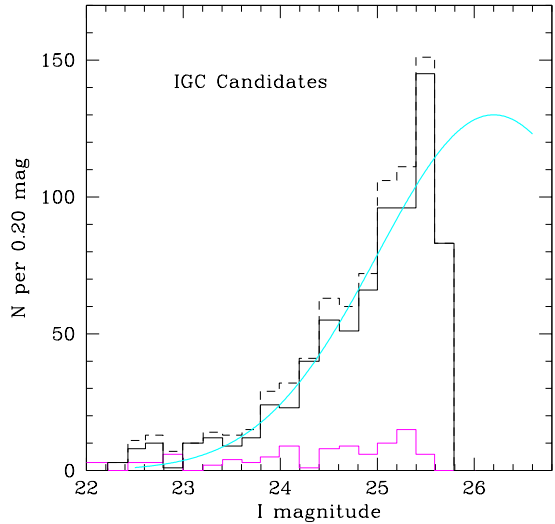}
\end{center}
\vspace{-0.2cm}
\caption{Luminosity distribution of the GC candidates.  The \emph{dashed line} shows the
number per 0.2-mag F814W bin, corrected for incompleteness and extinction $A_I$, while the \emph{magenta} 
histogram shows the number per bin for the background contamination.  The \emph{solid histogram}
shows the difference.  The \emph{cyan} curve shows an illustrative Gaussian luminosity
function for $(I^0, \sigma) = (26.2, 1.2)$.}
\vspace{0.0cm}
\label{fig:gclf}
\end{figure}

\begin{figure}[t]
\vspace{-0.0cm}
\begin{center}
\includegraphics[width=0.5\textwidth]{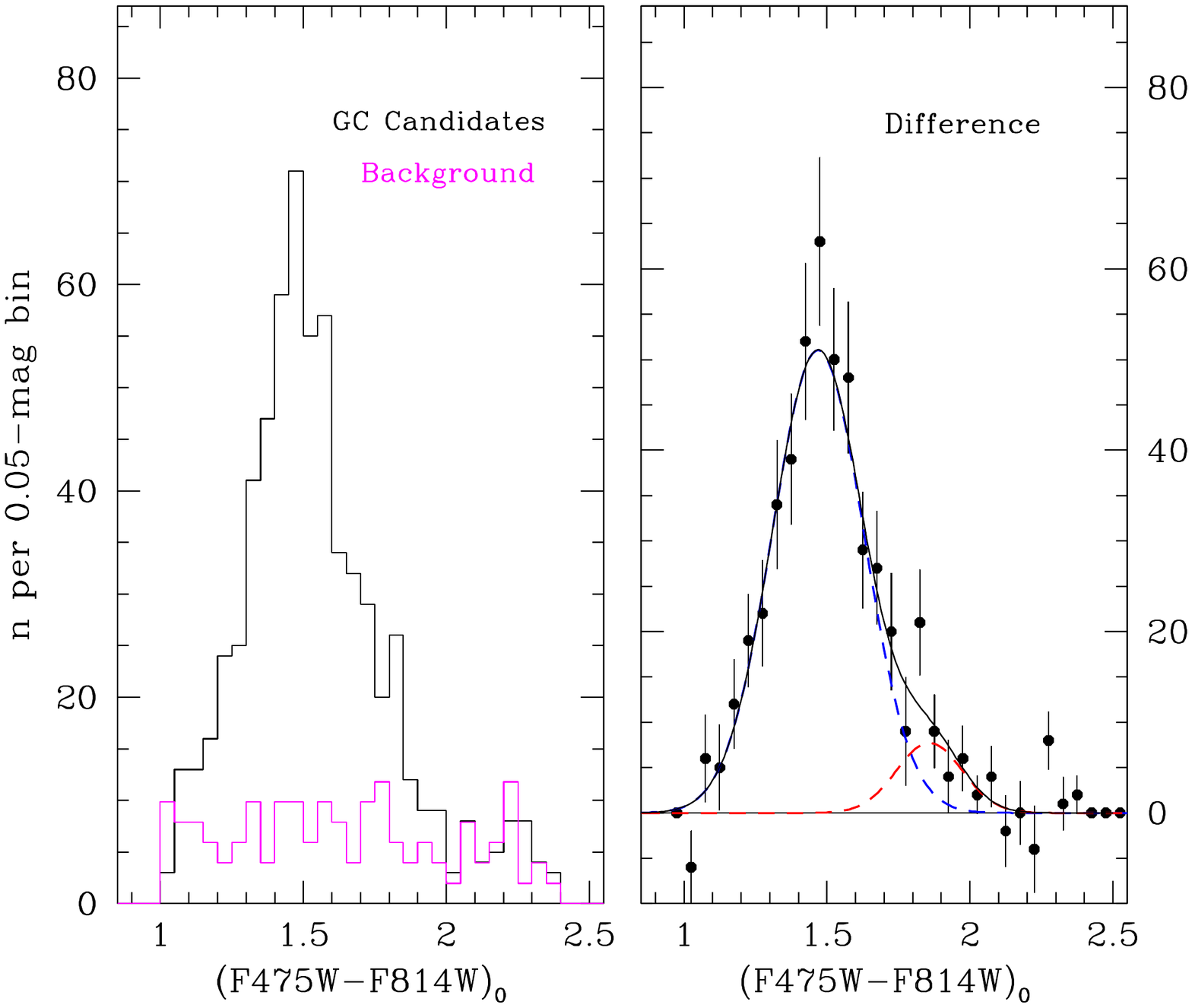}
\end{center}
\vspace{-0.2cm}
\caption{\emph{Left panel:} Color distribution histograms for the GC candidates in eight pairs
of outer fields (in black), and for objects in the six control fields (in magenta).  Number
of objects per 0.05-mag bin is plotted versus intrinsic color (F475W-F814W)$_0$.
The control-field data are normalized to the same area of 138 arcmin$^2$ as covered by the
Perseus fields minus V11/ACS (see text).
	\emph{Right panel:} Difference between the GC-candidate histogram and the control-field
histogram, giving a statistically ``clean'' color distribution function for the IGCs in Perseus.
Errorbars on each data point include the $N^{1/2}$ uncertainties in both GC candidates
and the background.
The bimodal-Gaussian fit to the distribution as described in the text is superimposed as the
black line, with the red and blue components shown as dashed lines.}
\vspace{0.0cm}
\label{fig:mdf}
\end{figure}

We have generated and measured the images for 
all of these control fields with exactly the same procedures as for our Perseus targets,  
including the same process of object selection and culling of nonstellar objects.
The images for the five HFF fields are taken in (F435W, F814W), and the images
for SS22A-PAR are taken in (F475W, F814W).  The HFF data were transformed to the
(F475W, F814W) system through
$$ F475W = F435W - 0.174 (F435W - F814W) $$
again derived from the equations from \cite{sirianni_etal2005}.
In Figure \ref{fig:cmdback_tile}, the results for the six control fields are shown
individually, while in Figure \ref{fig:cmdback}, all six are individually dereddened and combined.
This combined CMD represents a total area of 70 arcmin$^2$. 

In total, 93 objects fall clearly within the GC-candidate region defined above, making
an average of 15.5 per ACS field or equivalently 10 per WFC3 field.  
All of the six fields, however, are at higher Galactic latitude than is Perseus,
and in principle the numbers of contaminating objects falling within the GC
candidate selection box could be a function of latitude if they are dominated
by Galactic foreground stars rather than faint background galaxies.
An empirical test is shown in Figure \ref{fig:nbox}, 
where the numbers per field are plotted versus the absolute
value of Galactic latitude. Notably, the numbers do not depend significantly on $b$.
indicating that the contaminants within our magnitude and color range of interest 
are primarily from the Galactic halo rather than the disk.  
Over the entire 150 arcmin$^2$ area of our Perseus fields, these numbers would
therefore suggest that we should expect $\simeq 200$ contaminants in total
within the candidate box.  Averaging the 6 fields gives a mean
number density $\sigma_{bkgd} = (1.33 \pm 0.21)$ arcmin$^{-2}$.
The uncertainty of this mean value is about 1.5 times higher than expected 
from simple Poisson statistics, which we take to be due to intrinsic cosmic
scatter.  Substructure in the Milky Way halo, and clustering of faint background
galaxies over scales of a few arcminutes, will both contribute to this increased
variance. 

As an additional check on this estimate, we use the TRILEGAL
population model of the Milky Way \citep{girardi_etal2005} to gauge the
numbers of foreground stars at the latitude and longitude of Perseus.
These results are also shown in Figure \ref{fig:cmdback}
for a projected area on the sky of 150 arcmin$^2$, equal to the total
area of our (8+8) Perseus fields.  In this model, 143 stars fall within
the GC candidate box,
which makes up roughly 70\% of the total from the control fields given above.  
At face value, the TRILEGAL counts therefore 
suggest that only $\simeq30$\% of the control-field counts are from faint background
galaxies small enough to pass through our selection filters.

Perhaps more importantly, the distribution of model stars in Fig.~\ref{fig:cmdback}
is useful for showing where the CMD is most affected by contamination, which is the large
swath of points at upper right.  These are largely faint lower main sequence stars at 
various distances in the Milky Way disk (and to a lesser extent the halo), and their total numbers are highly
sensitive to Galactic latitude.  Fortunately, the Perseus GC candidates lie bluer and
fainter than these stars, in a region of the CMD where very little contamination
is present.

Within the GC candidate box, there are a total of 824 objects after rejection of
ones very near the obvious dwarf galaxies, as noted above.  In the same box, the data
from the control fields predict that $\sim 200$ of these should be ``background", for
a contamination rate of 24\%.  However, the background objects are quite evenly
distributed in color across the box (Fig.~\ref{fig:cmdback}) whereas the great majority
of the GC candidates are in the color range $1.0 < (F475W-F814W)_0 < 2.0$.  In this
narrower color range, there are 775 candidates but an estimated 156 background objects
in the same 150 arcmin$^2$ area, for a somewhat lower contamination fraction of 20\%.

These results indicate that the
contamination level in our sample of outer fields is small though not negligible.  
It is clear, however, that sparsely spread as they are,
IGCs are present and clearly measurable in these outlying Perseus fields.

\section{Results:  Characterizing the IGCs}

The HST data give high enough resolution, depth, and precision to
permit measurement of several characteristics of the IGCs and to make a series
of tests of how consistent their properties are compared with those in galaxies.
These include their scale sizes (half-light radii); their distribution in luminosity
(the GCLF); and their metallicity distribution as represented by color indices.
All of these results are consistent with the conclusion that the IGCs are
a generally normal population of GCs.

\begin{figure*}[t]
\vspace{-0.0cm}
\begin{center}
\includegraphics[width=0.7\textwidth]{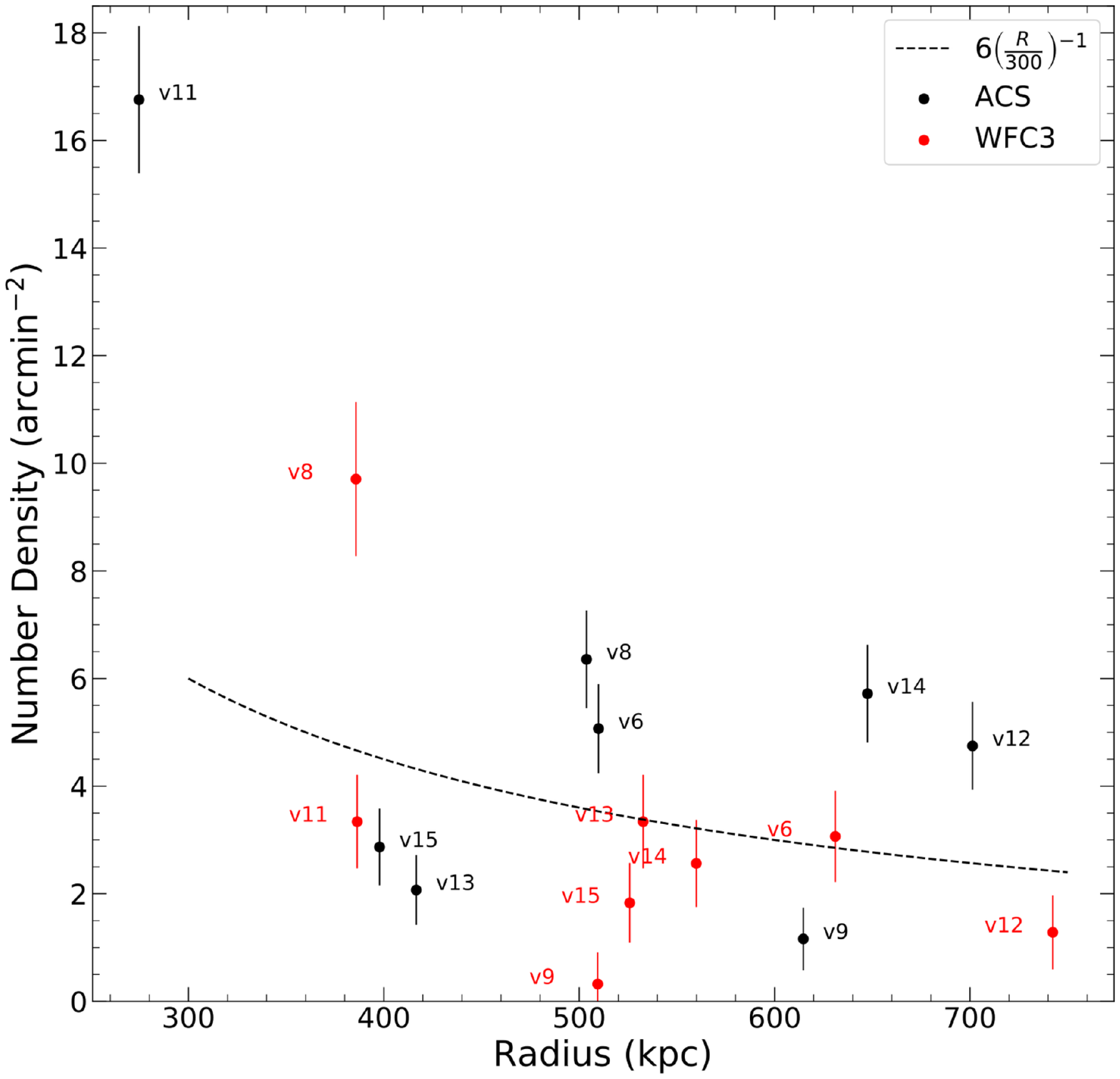}
\end{center}
\vspace{-5.0cm}
\caption{Number density of IGCs in the outer Perseus fields, plotted versus radius from
the Perseus center (assumed to be at NGC 1275).  Black symbols are ACS fields
and red symbols are the WFC3 fields.  The magnitude range included here is
$22.0 < F814W_0 < 25.5$.  The adopted mean backgrounddensity of 1.33 arcmin$^{-2}$
has been subtracted from all points.  Note V11 ACS (upper left) is
heavily contaminated by a nearby early-type galaxy.  The \emph{dashed line}
shows the $\phi \sim R^{-1}$ radial profile described in the text.
}
\vspace{0.9cm}
\label{fig:radial}
\end{figure*}

\subsection{Effective Radii}

Although by definition all the objects in the GC candidate list 
are reasonably well matched to the PSF shape within the specified 
range $|sharp| < 0.3$ as discussed above, estimates of 
intrinsic radii can be obtained by a more detailed fit of the PSF to each object.
We employ the code \emph{ishape} \citep{larsen1999} for this purpose, following many previous
instances of its use in the literature for GCs in remote galaxies
\citep[e.g.][among others]{larsen1999,larsen_etal2001,georgiev_etal2008,harris2009a,lim_etal2013}.  

To run \emph{ishape} we assume an intrinsic GC profile with central concentration
$c = r_t/r_c = 30$ or log $c$ = 1.5 (KING30 in the code notation), which is an average 
value for GCs \citep[][2010 edition]{harris1996}.  This assumed profile is then convolved 
with the individual PSF for each field and matched to each object, varying the assumed
{\sl fwhm} until a best fit is achieved.  We run \emph{ishape} only on the $I$-band images
because of their higher SNR.  

The HST resolution of {\sl fwhm} $= 0.1"$ is equivalent to 36 pc in linear scale.  Conservatively,
\emph{ishape} can measureably resolve a near-starlike object if its {\sl fwhm} is larger
than about 10\% of the stellar {\sl fwhm} \citep[see][]{larsen1999,harris2009a}.
For a KING30 model profile, the half-light radius is $r_h = 1.48 fwhm$.  
At the Perseus distance $d = 75$ Mpc, this resolvable limit then corresponds to $r_h(lim) \simeq 5$ pc.
In the Milky Way, the GC $r_h$ distribution peaks near 3 pc and almost none are
smaller than $r_h = 1$ pc, but the distribution has a long tail
to larger radii extending beyond 10 pc. 
Thus we should expect that many GCs in our candidate list will remain starlike,
but also that the larger GCs will be measurably resolved.
As a direct consistency test, \emph{ishape} was also applied to the artificial stars
used in the completeness tests described above. In principle, for these the fitted \emph{fwhm}
estimates should all be zero with some stochastically generated scatter.  

In Figure \ref{fig:histo} the $r_h$ measurements for 
all fields are displayed in histogram form.
The distribution of Perseus IGCs is clearly more extended than the histogram for
the artificial stars, which by definition are completely unresolved; 97\% of these
stars fall within the first bin ($r_h < 1$ pc).  Many of the GC candidates also
fall within $r_h < 1$ pc:  these should be a combination of genuine stars from
field contamination (see discussion above), plus genuine but small GCs.  

Notably, in the `resolved' range $r_h > 5$ pc, the shape of the Perseus histogram
resembles the high$-r_h$ tail of the Milky Way distribution also shown in Fig.~\ref{fig:histo}.  
Although the scale sizes of GCs occupy a large range from $\lesssim 1$ pc up to 10 pc
and more, their average size increases
with galactocentric distance \citep{gomez_woodley2007,harris2009a,harris_etal2010},
due to weaker galactic tidal-field limits at larger distance.  When this trend is coupled
with the prediction that IGCs are clusters stripped away from galactic halos
during interactions, IGCs should then be expected to have larger scale sizes than
the bulk of the GC population in the inner regions of galaxies.  In the Milky Way,
for example, the median size for 95 GCs lying within $R_{gc} < 8$ kpc is $r_h(med) = 2.5$ pc,
while for 59 clusters with $R_{gc} > 8$ kpc, $r_h(med) = 4.7$ pc, almost twice as large as for
the inner halo.  To the extent allowed by the resolution
limits of our imaging, we conclude that the scale-size distribution for
the Perseus candidates is quite consistent with their identification as GCs.

Intercomparisons of scale sizes ($r_h$) of GCs within other galaxies
have been carried out for numerous galaxies in Virgo, Fornax, and other relatively
nearby environments \citep{jordan_etal05,masters_etal2010,harris2009a,gomez_woodley2007}.
These studies show close consistency of the $r_h$ distributions between galaxies, 
including the Milky Way itself
\citep[enough so that the $r_h$ distribution can be used as a rough distance indicator; see][]{jordan_etal05}.  Instead, we would like in principle to compare the distribution in 
Fig.~\ref{fig:histo} with other IGC populations.  The only such high-quality data available are
the scale size measurements for just four IGCs in Virgo measured from HST imaging \citep{williams_etal2007} 
(though see also \citet{ko_etal2017} for two other IGCs measured from ground-based imaging):
these four have $r_h = 2.0, 3.4, 8.0,$ and 9.6 pc. All of these fit comfortably within
the Perseus IGC range, and their average $\langle r_h \rangle = 5.8$ pc is near the
peak of the Perseus IGC distribution.

\subsection{Luminosity Function}

Old GC populations consistently follow a Gaussian-like
luminosity function (LF) in number per unit magnitude \citep[e.g.][]{jordan2007,villegas2010,harris_etal2014},
so an additional test of the nature of the Perseus candidates is the shape of
their LF.  This is shown in Figure \ref{fig:gclf} for the combined candidate sample.
At present, only a rough test can be made 
since the photometric limit falls short of the normal GCLF peak frequency (turnover point).
The previous work of HM17 shows similar results,
though their data are mostly from the Perseus core galaxies.

The turnover (peak) absolute magnitude of the GCLF, and its dispersion $\sigma$ have
typical values $M_I^0 \simeq -8.4$ and $\sigma \simeq 1.2$ for galaxies comparable
to the Milky Way.  However, 
both the turnover and dispersion depend systematically on host galaxy luminosity 
\citep[cf.][]{villegas2010} such that larger galaxies have
progressively more luminous turnovers (at the rate of $\simeq 0.04$ mag per galaxy 
absolute magnitude) and broader GCLFs (at the rate of $\simeq 0.1$ mag per galaxy
absolute magnitude).  Recent modelling indicates that the
biggest contributions to the ICL are stars stripped from dwarf-sized to intermediate-sized galaxies
\citep{kharris_etal2017,ramos_etal2018}. Such galaxies are 3 mag or more
fainter than the giants that much of the previous GCLF literature has concentrated
on.  Thus in principle, accurate determination of the GCLF parameters for the IGC component
would provide an additional observational test of its origin.  
However, if the photometric limit falls short of the GCLF turnover as is the case here,
both $I^0$ and $\sigma$ cannot
be reliably solved simultaneously \citep[e.g.][]{hanes_whittaker1987}, and for the present
we are limited to a simple consistency test.
In Fig.~\ref{fig:gclf} an illustrative curve is shown for a Milky-Way-like system
with turnover $I^0 = 26.2$ and dispersion $\sigma = 1.2$.
Again, the shape of the GCLF after subtraction of the background LF is consistent
in broad terms with a normal GC population.

At the bright end of the GCLF ($I < 23.5$) a slight excess number of objects is present
above the fitted Gaussian curve.  Some of these objects could be UCD candidates,
which are known to be present in Perseus in significant numbers \citep{penny_etal2011,penny_etal2012}.
These are visible in the CMD of Fig.~\ref{fig:cmdall} as a scattering of points above
the main GC population.  Similar small excess numbers of such objects can be seen in the
GCLFs of major galaxies \citep{harris_etal2014}, and with no additional information to
go on, it is difficult to distinguish between very luminous GCs and lower-luminosity UCDs. 

\subsection{Color Distribution: Blue and Red Fractions}

In most luminous galaxies the color-index distribution function (CDF) for GCs is roughly bimodal, with 
``blue'' and ``red'' sequences separated by $\sim 1$~dex in metallicity [Fe/H]
\citep[e.g.][]{larsen_etal2001,brodie_strader06,peng_etal2006,harris2009a,harris_etal2017a}.  
This is not a universal phenomenon, however,
and for the most luminous giants in particular (the Brightest Cluster Galaxies or BCGs), 
the CDF becomes more uniformly populated
\citep{harris_etal2016,harris_etal2017a}.
Models for the generation of the ICL and IGCs (cf. the references cited above) further demonstrate
that the higher-metallicity GCs should follow a spatial distribution more centrally concentrated
to the cluster core, like the member galaxies, whereas the low-metallicity GCs stripped primarily
from the dwarfs and outer halos of the member galaxies should follow a more extended distribution
closer to the dark-matter potential well.  The IGC data from across the Virgo cluster
\citep{durrell_etal2014} and Coma \citep{peng_etal2011,madrid_etal2018} are consistent with this prediction.

The CDF from our pairs of outer fields
is shown in Figure \ref{fig:mdf} for all the objects within the
GC candidate box, plotted as number versus intrinsic color and in $0.05-$mag color bins.
Here, we deliberately exclude the data from V11 ACS, because of its proximity to
a large galaxy and its obviously higher number density of candidates.
The CDF is also shown for the total numbers of objects in the six control fields,
now normalized to the 138 arcmin$^2$ area covered by the (7 ACS + 8 WFC3) Perseus fields. 

The GC candidates show a major, strong peak near (F475W-F814W)$_0 \simeq 1.5$, which
can be identified as the metal-poor GC population.  There is a hint of a redder
component but it is clearly minor.  By comparison,
the CDF for the control fields is quite uniform versus color, showing no obvious peaks.
Another noteworthy feature is that for (F475W-F814W)$_0 \gtrsim 2.0$ the numbers of
background objects almost precisely match up with the GC candidates.

The difference histogram (GC candidates \emph{minus} background) is shown in the right panel
of Fig.~\ref{fig:mdf}.  The very reddest objects
with $(F475W-F814W)_0 > 2.1$  have subtracted out well, leaving a statistically
pure sample of GCs whose CDF can be matched to a conventional bimodal-Gaussian model
to assess the relative fractions of metal-poor or metal-rich clusters.
Of the various fitting codes in the recent literature, 
we use RMIX \citep[e.g.][]{wehner_etal2008}, which can handle 
a user-prepared histogram of the CDF.
The more frequently used KMM or 
GMM fitting code \citep{muratov_gnedin2010}, for example, assumes a clean, uncontaminated
list of objects (i.e., GCs only) and therefore is less suitable for a dataset containing 
significant contamination.

RMIX shows that a Gaussian \emph{unimodal} fit is strongly rejected, at $>$99\% 
significance.  For a bimodal fit,
the results give best-fit components at colors
$\mu_1 = 1.469 \pm 0.013, \mu_2 = 1.858 \pm 0.048$; and
dispersions $\sigma_1 = 0.160 \pm 0.009, \sigma_2 = 0.115 \pm 0.025$.    
The blue component makes up a fraction $f_{blue} = 0.902 \pm 0.036$ of the total, 
thus giving $f_{red} = 0.098$.  The total fit, shown in Fig.~\ref{fig:mdf},
accurately matches the CDF with no clear evidence that any other components are
needed within the uncertainties of the data.

We conclude that at least in the outer regions of Perseus, the IGCs are strongly dominated
by metal-poor clusters (making up 90\% of the total).  Small though it is, however,
the red component is significant; the total CDF would not be well fitted without it.
A first-order conversion of the color indices to approximate metallicity differences can be made
if we combine $\Delta(B-I) = 1.222 \Delta (F475W-F814W)$ with
$\Delta(B-I) \simeq 0.375 \Delta{\rm [Fe/H]}$ \citep{harris2009a}, giving
$\Delta$[Fe/H] $\simeq 3.26 \Delta (F475W-F814W)$.  The color difference
$\mu_2-\mu_1 = (0.389 \pm 0.050)$ thus corresponds to $\Delta$[Fe/H] $\simeq 1.2 \pm 0.1$,
similar to what is found in individual large galaxies 
with well populated MDFs \citep[e.g.][]{harris2009a,brodie_strader06}.
The dispersion of the blue mode $\sigma_1$ corresponds to $\sigma_1$[Fe/H] $\simeq 0.49 \pm 0.03$,
while the red mode has $\sigma_2$[Fe/H] $\simeq 0.37 \pm 0.07$.  These dispersions are
larger than those in the Milky Way \citep{harris_etal2016}, which are
$\sigma_1 = 0.38$ dex, $\sigma_2 = 0.23$ dex.  Part of the difference is likely to be
due simply to photometric scatter, but part of it might well be a result of the origin of the IGCs
as a composite population drawn from many individual galaxies.

Tracing out where the redder IGCs are located in the various fields shows no detectable
concentration towards the known galaxies in the fields.  Admittedly, however, such tests are
limited by the very small numbers of the red GCs.


We defer additional discussion of
this interesting direction till more evidence is available
from the CDFs of the major core galaxies. But for the
present we suggest that the observed CDF is consistent with a 
normal range of GC metallicities.

\subsection{Radial Distribution}

The eight target pairs covering the Perseus IGM that we discuss here have a wide
range of radii from the center of Perseus (adopted as the position of NGC 1275, the
central cD galaxy).  We can therefore use these to gain a preliminary idea of the
projected density of the IGC population versus radius.  
The present results are shown in Figure \ref{fig:radial}.  Here, the numbers of 
objects in the GC selection box, after subtraction of the background density level
$\phi_{bkgd} = 1.33$ arcmin$^{-2}$,
are plotted as number density versus radius within Perseus, where NGC 1275
is assumed to be at the Perseus center.  

For comparison, the outermost fields discussed by HM17 were the shallow WFPC2 images
at $R \sim 440$ kpc, so our present dataset extends almost twice as far out.
In Fig.~\ref{fig:radial}, no strong radial dependence is evident; instead, the
dominant impression is the field-to-field scatter.  HM17 estimated that the IGCs
follow a projected density profile $\phi \sim R^{-1.0 \pm 0.4}$, which is
also shown in Fig.~\ref{fig:radial}.  Within the scatter, 
a falloff with radius this shallow is consistent with the data and would also be
consistent with the shape of the isothermal dark-matter potential well of Perseus as a whole.

If the field-to-field scatter that is quite noticeable 
in Figs.~\ref{fig:xyall} and \ref{fig:radial} is not due 
strictly to simple stochastic differences in an already-low mean density, part of the
explanation may lie in the particular locations of these fields, which are not
randomly located across the Perseus region. 
V12/ACS, V13, and V14/ACS fall along a prominent chain 
of galaxies running from the Perseus core region (East side) through to IC 310  
(the giant ETG on the West side, at lower right in Fig.~\ref{fig:fields}) and beyond.  
This connection raises the possibility that IGCs may also lie preferentially along
the same axis.
On a scale an order of magnitude larger, the distribution of Perseus galaxies connects
with the Perseus-Pisces supercluster \citep{haynes_giovanelli1986,wegner_etal1993}.
V11/ACS, V8/ACS, and V15/ACS lie moderately close to major galaxies to the south and northwest.
Similarly, the pointings with the lowest number densities (V9, V15/WFC3, V12/WFC3)
lie furthest from identifiable substructures. 
In upcoming work, we will use the Subaru HSC imaging material to gain a more
comprehensive picture of the IGCs, including especially their global
spatial distribution and their total numbers.

Lastly, it is worth noting that the redder IGCs, though small in number, can
be found in even the target fields furthest from the Perseus center.  For comparison,
\citet{longobardi_etal2018} found a similar result for IGCs in the Virgo cluster:
The bluer, lower-metallicity IGCs define a slightly shallower radial distribution than
do the higher-metallicity ones, but both components are present to the outermost
radii surveyed.


\section{Summary}

We introduce a new imaging program (PIPER) for the giant Perseus cluster of galaxies.
The main goals for this program are to survey the globular cluster populations
around the giant galaxies in the Perseus core, in 40 UDG candidates, and in the IntraCluster Medium,
in addition to building a more comprehensive sample of the UCDs and cEs in Perseus.
Upon completion, the program will include photometry in 35 different HST pointings
extending from the cluster
center (NGC 1275) out to projected radial distances more than 700 kpc from center,
as well as Subaru HSC multicolor photometry covering the entire Perseus region
to a similar radius.

In the present paper, we outline the photometric reduction methodology for our HST
images, and describe the first round of results obtained from eight ACS/WFC3 image pairs
in the (F475W, F814W) filters (for ACS) and (F475X, F814W) (for WFC3).
These HST exposures are designed to be deep enough to 
approach the LF turnover point of the globular cluster systems in the Perseus galaxies.
The eight pairs of fields analyzed here are all located well outside the Perseus
core, and we use these to find and characterize the intergalactic globular
cluster population (IGC).
Field contamination is determined from six Archival HST/ACS fields of similar depth and with
similar filters.

The results of this first PIPER study are as follows:
\begin{enumerate}
\item The dereddened photometry is used to define a ``GC candidate'' region in the color-magnitude diagram 
with boundaries $22.0 < F814W_0 < 25.5$, $1.0 < (F475W-F814W)_0 < 2.4$.
Even at distances $R > 700$ kpc, a sparse but clearly
detectable population of objects is present that matches the color and magnitude ranges
expected for normal GCs. In almost all of our fields
these IGCs dominate over the field contamination.
\item The half-light radii for GCs larger than $r_h \gtrsim 5$ pc can be measured
(smaller ones are unresolved even with HST).
For these, the size distribution resembles that of the Milky Way GCs.  A peak
near $r_h \simeq 5$ pc is followed by a long tail extending up to $r_h \sim 15$ pc.
\item The luminosity distribution has the characteristic Gaussian shape in number
per unit magnitude, with an estimated turnover point at $I_0 \simeq 26.2$ ($M_I = -8.4$)
and dispersion $\sigma \simeq 1.2$ mag.  
\item The color distribution is measurably bimodal, but with   
90\% of the IGCs belonging to the blue (metal-poor) mode.  Conversely, even in these
very remote fields a trace population of red (metal-rich) clusters is present,
indicating that at least some of the IGC population must have originated in
moderately large galaxies.
\item Finally, the distribution of the IGCs with radius from the center of the
Perseus cluster is consistent with a shallow $R^{-1}$ falloff, though this
trend is obscured by field-to-field scatter.
\end{enumerate}

In future papers of this series, analyses will concentrate on the UDGs and their own GC populations,
the central giant galaxies, and the UCDs and compact ellipticals to be discovered in our target fields.
Adding in the HSC data from Subaru will also give us a more comprehensive assessment of the
IGC spatial distribution on its broadest scales.

\acknowledgements

WEH acknowledges financial support 
from the Natural Sciences and Engineering Research Council of Canada.
AJR was supported by National Science Foundation grants AST-1515084
and 1616710 and as a
Research Corporation for Science Advancement Cottrell Scholar.  
AJR and PRD are supported by funds provided by NASA through grant GO-15235
from the Space Telescope Science Institute, which is operated by the
Association of Universities for Research in Astronomy, Incorporated,
under NASA contract NAS5-26555.
JB acknowledges support from NSF grants AST-1616598 and AST-1518294.
CW and TL were supported by the Deutsche Forschungsgemeinschaft (DFG, German Research Foundation) 
through project 394551440.
TL also acknowledges financial support from the European Union's Horizon  
2020 research and innovation programme under the Marie Sk\l{}odowska-Curie grant agreement 
No.\ 721463 to the SUNDIAL ITN  network.
Lastly, we are grateful to the referee for a very constructive review.

\vspace{5mm}
\facility{{\sl HST} (ACS, WFC3)}

\makeatletter\@chicagotrue\makeatother
\bibliographystyle{apj}
\bibliography{gc}

\label{lastpage}

\end{document}